\documentclass[12pt]{article}
\pdfoutput=1
\usepackage{graphicx}
\usepackage{epstopdf}
\usepackage{amsmath}
\usepackage{amsfonts}
\usepackage{amssymb}
\usepackage{color}
\usepackage{mathrsfs}
\usepackage[utf8]{inputenc}
\usepackage{subcaption}
\usepackage{array}
\usepackage{booktabs}
\usepackage{tabularx}
\usepackage{changepage}
\usepackage{lipsum}

\usepackage[colorinlistoftodos]{todonotes}
\usepackage[font={footnotesize}]{caption}

\newlength{\offsetpage}
\newenvironment{widepage}{\begin{adjustwidth}{-\offsetpage}{-\offsetpage}%
    \addtolength{\textwidth}{2\offsetpage}}%
{\end{adjustwidth}}

\pdfoutput=1

\usepackage[english]{babel}

\usepackage[colorlinks,bookmarks]{hyperref}
\definecolor{linkblue}{rgb}{0,0,0.8}
\definecolor{linkgreen}{rgb}{0,0.5,0}

\hypersetup{pdfpagemode=UseNone, pdfstartview=FitH, linkcolor=linkblue, %
            citecolor=linkgreen, urlcolor=linkblue}

\allowdisplaybreaks

\numberwithin{equation}{section}

\newcommand{\bea}{\begin{eqnarray}}
\newcommand{\eea}{\end{eqnarray}}
\newcommand{\be}{\begin{equation}}
\newcommand{\ee}{\end{equation}}

\newcommand{\n}{\nonumber \\}

\newcommand{\knl}{k_{\rm NL}}

\newcommand{\eqn}[1]{Eq.~(\ref{#1})}

\newcommand{\Comment}[1]{{}}
\newcolumntype{P}[1]{>{\centering\arraybackslash}p{#1}}

\begin{document}
\def\thefootnote{\fnsymbol{footnote}}

\setcounter{page}{1} \baselineskip=15.5pt \thispagestyle{empty}

\begin{flushright}
\end{flushright}

\begin{center}

{\Large \bf On the Bispectra of Very Massive Tracers in the Effective Field Theory of Large-Scale Structure\\[0.7cm]}
{\large  Ethan O.\ Nadler$^{1}$, Ashley Perko$^{1,2,3,4}$ and Leonardo Senatore$^{1,2,3}$}%\\[0.3cm] Leonardo Senatore$^{1,2,3}$}
\\[0.7cm]
%\vspace{.7cm}
%\vspace{.3cm}

{\normalsize { \sl $^{1}$ Kavli Institute for Particle Astrophysics and Cosmology and Department of Physics, Stanford University, Stanford, CA 94305, USA}}\\
\vspace{.3cm}

{\normalsize { \sl $^{2}$ Stanford Institute for Theoretical Physics,\\ Stanford University, Stanford, CA 94306, USA}}\\
\vspace{.3cm}

{\normalsize { \sl $^{3}$ SLAC National Accelerator Laboratory, \\ Menlo Park, CA 94025, USA}}\\
\vspace{.3cm}

{\normalsize { \sl $^{4}$ Department of Physics and Astronomy, Dartmouth College,\\ 6127 Wilder Laboratory, Hanover, NH 03755, USA}}\\
\vspace{.3cm}

\end{center}

\vspace{.8cm}

%%%%%%%%%%%%%%%%%%
%
%           abstract
%
%
%%%%%%%%%%%%%%%%%%

\hrule \vspace{0.3cm}
{\small  \noindent \textbf{Abstract} \\[0.3cm]

\noindent The Effective Field Theory of Large-Scale Structure (EFTofLSS) provides a consistent perturbative framework for describing the statistical distribution of cosmological large-scale structure. 
In a previous EFTofLSS calculation that involved the one-loop power spectra and tree-level bispectra, it was shown that the $k$-reach of the prediction for biased tracers is comparable for all investigated masses if suitable higher-derivative biases, which are less suppressed for more massive tracers, are added. However, it is possible that the non-linear biases grow faster with tracer mass than the linear bias, implying that loop contributions could be the leading correction to the bispectra. To check this, we include the one-loop contributions in a fit to numerical data in the limit of strongly enhanced higher-order biases. We show that the resulting one-loop power spectra and higher-derivative plus leading one-loop bispectra fit the two- and three-point functions respectively up to $k\simeq 0.19\ h\ \rm{Mpc}^{-1}$ and $k\simeq 0.14\ h\ \rm{Mpc}^{-1}$ at the percent level. We find that the higher-order bias coefficients are not strongly enhanced, and we argue that the gain in perturbative reach due to the leading one-loop contributions to the bispectra is relatively small. Thus, we conclude that higher-derivative biases provide the leading correction to the bispectra for tracers of a very wide range of masses.

\vspace{0.3cm}
\hrule

\def\thefootnote{\arabic{footnote}}
\setcounter{footnote}{0}

%%%%%%%%%%%%%%%%%%
%
%
%     Introduction
%
%
%%%%%%%%%%%%%%%%%%
\newpage
\tableofcontents
\section{Introduction}

Analytic descriptions of the statistical distribution of cosmological large-scale structure will be crucial in order to analyze the plethora of data from upcoming galaxy surveys such as The Dark Energy Spectroscopic Instrument \cite{2013arXiv1308.0847L} and the Large Synoptic Survey Telescope \cite{2009arXiv0912.0201L}. To this end, it is essential to understand the perturbative reach of our theoretical predictions in order to exploit as much of the forthcoming survey data as possible.

The Effective Field Theory of Large-Scale Structure (EFTofLSS) \cite{Baumann:2010tm,Carrasco:2012cv,Porto:2013qua,Senatore:2014via} provides an analytic framework for predicting LSS correlation functions in the mildly non-linear regime \cite{Baumann:2010tm,Carrasco:2012cv,Porto:2013qua,Senatore:2014via,Carrasco:2013sva,Carrasco:2013mua,Pajer:2013jj,Carroll:2013oxa,Mercolli:2013bsa,Angulo:2014tfa,Baldauf:2014qfa,Senatore:2014eva,Senatore:2014vja,Lewandowski:2014rca,Mirbabayi:2014zca,Foreman:2015uva,Angulo:2015eqa,McQuinn:2015tva,Assassi:2015jqa,Baldauf:2015tla,Fujita:2016dne,Perko:2016puo,Lewandowski:2016yce,Lewandowski:2017kes,Senatore:2017hyk,Senatore:2017pbn}. Recently, the EFTofLSS has been used to describe the clustering of biased tracers of the dark matter distribution \cite{Senatore:2014eva}, and the EFTofLSS prediction for the one-loop power spectra and tree-level bispectra of low-mass tracers has been shown to fit two- and three-point functions from numerical simulations within a few percent up to $k\simeq 0.17\ h\ \rm{Mpc}^{-1}$ \cite{Angulo:2015eqa,Fujita:2016dne}\footnote{We note that the simulation data for the three-point statistics of very massive tracers used in \cite{Angulo:2015eqa} and~\cite{Fujita:2016dne} extend only to $k = 0.14\ h\ \rm{Mpc}^{-1}$, and we use the same data in our analysis.}. However, the perturbative reach of the tree-level EFTofLSS prediction for the bispectra of biased tracers decreases for more massive objects. In particular, the prediction for the one-loop power spectra and tree-level bispectra of very massive tracers fails to fit simulation data at $k\simeq 0.12\ h\ \rm{Mpc}^{-1}$ \cite{Angulo:2015eqa}. %\footnote{The EFTofLSS prediction fits the halo-halo and halo-matter power spectra alone up to $k\simeq 0.30\ h\ \rm{Mpc}^{-1}$ at the percent level, independent of tracer mass \cite{Angulo:2015eqa}.}. 
\cite{Fujita:2016dne} argues that the perturbative reach of this prediction decreases with tracer mass because of the potentially large higher-derivative biases that are not included in the tree-level bispectra. Indeed, correlation functions of biased tracers in the EFTofLSS are expressed as sums of correlation functions of dark matter fields weighted by bias coefficients. While higher-derivative and perturbative contributions to dark-matter correlation functions are expanded in the same parameter for all tracers (all proportional to powers of $k/\knl$, where $\knl$ is the wavenumber associated to the non-linear scale), higher derivative biases scale as powers of $(k/k_{\rm M})^2$, where $k_{\rm M}$ is the wavenumber associated to the mass scale of the tracer, which is lower for more massive tracers. By adding higher-derivative biases to the halo-matter-matter, halo-halo-matter, and halo-halo-halo bispectra, \cite{Fujita:2016dne} is able to fit the two- and three-point functions from the Millennium-XXL N-body simulation \cite{Angulo12033216} for tracers of \emph{all} masses up to $k\simeq 0.17\ h\ \rm{Mpc}^{-1}$ at the percent level. Thus, \cite{Fujita:2016dne} concludes that it is necessary to include higher-derivative biases in a consistent calculation of the bispectra for very massive tracers, and that doing so ensures a $k$-reach of $k\simeq 0.17\ h\ \rm{Mpc}^{-1}$ for the two- and three-point statistics of all collapsed objects investigated in their analysis.

The authors of \cite{Fujita:2016dne} note that one-loop contributions to the bispectra for very massive tracers are small relative to higher-derivative biases provided that higher-order bias coefficients do not grow rapidly with respect to the linear bias as tracer mass increases\footnote{We refer to tracers according to the mass bins assigned in \cite{Fujita:2016dne}. The lightest tracers (``Bin 0'') roughly correspond to $10^{12}\ \rm{M}_{\rm \odot}$ halos, and the most massive tracers (``Bin 3'') correspond to $10^{14}\ \rm{M}_{\rm \odot}$ halos.}. However, it is not necessarily accurate to assume that the ratio of the higher-order bias coefficients to the linear bias depends weakly on tracer mass, and doing so might influence the resulting fit. In particular, if the bias coefficients scale as $b_n\sim b_1^{n}$, where $b_n$ denotes an $n$-th order bias coefficient, then the fit with the tree-level bispectra for very massive tracers would fail due to a lack of certain one-loop contributions rather than a lack of higher-derivative biases. It is therefore important to check whether these one-loop contributions should be included alongside higher-derivative biases in a consistent calculation of the bispectra for very massive tracers. For the purposes of this check, it is sufficient to adopt the following strategy. First, we \emph{assume} that $b_n\sim b_1^{n}$, which we refer to as the high-bias-scaling limit. We will show that it is only necessary to include a small, easy-to-compute subset of the full one-loop bispectra for biased tracers in this limit, and we calculate these diagrams below. We find that adding these one-loop contributions to the tree-level plus higher-derivative bispectra improves the $k$-reach of the fit to the two- and three-point statistics of very massive tracers to $k\simeq 0.19\ h\ \rm{Mpc}^{-1}$, which is only a marginal improvement over the results in \cite{Fujita:2016dne}. Moreover, we show that these one-loop contributions improve the fit only slightly more for very massive tracers than for light tracers, and we do \emph{not} find strong evidence that the best-fit bias coefficients obey $b_n\sim b_1^{n}$. Thus, we conclude that the calculation in~\cite{Fujita:2016dne} is consistent.

Nevertheless, including the aforementioned one-loop contributions increases the functional freedom of the prediction for the bispectra. Specifically, adding these contributions introduces new bias coefficients and breaks certain parameter degeneracies, resulting in four additional bias coefficients relative to the calculation in \cite{Fujita:2016dne}. Furthermore, the natural size of the one-loop contributions is not completely negligible at high wavenumbers. As a result, both the $k$-reach of the prediction for very massive tracers and the overall goodness of fit at lower wavenumbers are improved when we include these one-loop contributions. This is consistent with the fact that the overall fit to the power spectra and bispectra for very massive tracers in \cite{Fujita:2016dne}, which includes only higher-derivative corrections to the bispectra, slightly underperforms relative to the fit for low-mass tracers. In particular, \cite{Fujita:2016dne} somewhat arbitrarily increases the best-fitting wavenumber that enters the $p$-value calculation for their Bin 3 fit in order to achieve a $k$-reach of $k\simeq 0.17\ h\ \rm{Mpc}^{-1}$.

Our calculation confirms that the EFTofLSS prediction presented in \cite{Fujita:2016dne} fits the power spectra and bispectra for tracers of a wide range of masses up to $k\simeq 0.17\ h\ \rm{Mpc}^{-1}$ without making spurious assumptions about the size of the one-loop contributions or the scaling of the higher-order bias coefficients. Moreover, we find that the leading one-loop contributions in the high-bias-scaling limit are small relative to the higher-derivative terms in practice, except at the highest scales of interest. We argue that this is an important check of the prediction in \cite{Fujita:2016dne}, since the fact that these potentially bias-enhanced one-loop contributions could be comparable to the higher-derivative terms means that the calculation in \cite{Fujita:2016dne} could have been inconsistent. Our results demonstrate the predictive power of the EFTofLSS: we conclude that the two-point functions (i.e., the halo-halo and halo-matter power spectra) and the three-point functions (i.e., the halo-halo-halo, halo-halo-matter, and halo-matter-matter bispectra) for biased tracers can be fit simultaneously up to $k\simeq 0.17\ h\ \rm{Mpc}^{-1}$ at the percent level using the seven- (for light tracers) to twelve- (for massive tracers) parameter prediction derived in \cite{Fujita:2016dne}. We reiterate, however, that simulated bispectra are only available up to $k = 0.14\ h\ \rm{Mpc}^{-1}$ in these fits.%Along with confirming this result, our work addresses the question of how to describe very massive objects in the EFTofLSS framework.

This paper is organized as follows. In \S\ref{sec:bispectrum}, we study the bispectra of very massive tracers in the EFTofLSS. In \S\ref{sec:corrections}, we consider corrections to the tree-level bispectra and we show that it is necessary to check whether the leading one-loop contributions in the high-bias-scaling limit should be included alongside higher-derivative biases. In \S\ref{sec:loop}, we calculate the one-loop contributions to the bispectra in the high-bias-scaling limit, and in \S\ref{sec:theory} we summarize the resulting EFTofLSS prediction for the two- and three-point functions of biased tracers\footnote{The codes used for the calculations in this paper are publicly available on the \href{http://stanford.edu/~senatore/}{EFTofLSS repository}.}. We describe our fits to simulation data in \S\ref{sec:fit}, and we present our results in \S\ref{sec:results}. We conclude in \S\ref{sec:conclusion}.

%----------------------------------------------------------------------------------------
%	SECTION 2
%----------------------------------------------------------------------------------------

\section{EFTofLSS Prediction for the Bispectra of Very Massive Tracers}
\label{sec:bispectrum}

\subsection{Corrections to the Tree-Level Bispectra}
\label{sec:corrections}

The EFTofLSS prediction for the one-loop power spectra and tree-level bispectra of low-mass tracers fits simulation data up to $k\simeq 0.17\ h\ \rm{Mpc}^{-1}$, but this prediction fails at $k\simeq 0.12\ h\ \rm{Mpc}^{-1}$ for very massive tracers. \cite{Fujita:2016dne} argues that this occurs because the expansion parameter $(k/k_{M})^{2}$ associated with higher-derivative biases grows with tracer mass, implying that higher-derivative biases must be included for very massive tracers. Here, $k_{M}$ is the wavenumber associated with a collapsed object of mass $M$, so we expect that $M \sim \rho/k_{M}^3$, where $\rho$ is the mean matter density in the universe. Thus, higher-derivative biases become increasingly important as tracer mass increases, which could explain why the perturbative reach of the tree-level prediction for the bispectra decreases with tracer mass. However, we must also consider the one-loop contributions to the bispectra that enter with powers of $ (b_n/b_1)(k/k_{\rm NL})^{3+n}$, where $k_{\rm NL}$ is the wavenumber corresponding to the nonlinear scale, $n \sim -1.7$ is the slope of the matter power spectrum, and $b_n$ stands for an $n$-th order bias coefficient. Since the tree-level bispectra do not include such contributions, these one-loop terms could also account for the relatively poor perturbative reach of the prediction for very massive tracers depending on how $b_n/b_1$ scales with tracer mass\footnote{We do not consider higher-order loops in this paper, since we find that the relevant one-loop contributions are generally small relative to the higher-derivative terms.}.

In \cite{Fujita:2016dne}, the authors estimate that the $(k/k_{M})^{2}$ factor associated with the higher-derivative biases grows more quickly with tracer mass than the $(k/k_{\rm NL})^{3+n}$ scaling of one-loop contributions. Thus, they argue that higher-derivative biases must be included in a consistent calculation of the bispectra for very massive tracers while one-loop contributions can be ignored. However, this argument does not account for the bias coefficients that enter with the one-loop contributions. In particular, it is possible that higher-order bias coefficients grow with some large power of $b_1$ as tracer mass increases. Indeed, for extremely massive tracers, recent measurements of the linear, quadratic, and cubic bias in N-body simulations indicate that $b_n \sim b_1^n$ (for example, see \cite{Lazeyras:2015lgp}); however, it is not clear whether this scaling applies for the less extreme high-mass tracers considered in \cite{Fujita:2016dne}\footnote{By inspecting the results in \cite{Lazeyras:2015lgp}, we infer that the scaling $b_n\sim b_1^n$ appears to be a good approximation for the relevant subset of bias coefficients measured in \cite{Lazeyras:2015lgp} only for very massive cluster-size halos. This will be confirmed by our fits, which indeed measure the full set of biases in the EFTofLSS that contribute at the order of our calculation.}. Nonetheless, since the higher-order bias coefficients can potentially grow steeply with tracer mass, it is possible that the fit with the tree-level bispectra for very massive tracers fails at relatively low wavenumbers because bias-enhanced loop terms, rather than $(k/k_{M})^{2}$-enhanced higher-derivative terms, are not included. The best-fit bias coefficients for the fits in \cite{Fujita:2016dne} do not suggest that $b_n/b_1$ grows with tracer mass, which would suggest that their calculation is consistent. However, including higher-derivative biases without including certain potentially-leading one-loop contributions could influence the resulting fit, since adding only higher-derivative terms implicitly assumes that higher-order bias coefficients are comparable in magnitude to $b_1$ (i.e., $b_n/b_1\sim 1$). Thus, it is important to check whether these potentially bias-enhanced one-loop contributions are significant. To do so, we \emph{assume} that the very massive tracers considered in \cite{Fujita:2016dne} have non-linear biases that are maximally enhanced, in the sense that $b_n\gg b_1$, which allows us to identify the leading subset of the one-loop contributions in this high-bias-scaling limit. By adding these one-loop contributions to the tree-level plus higher-derivative prediction for the bispectra of very massive tracers, we can check whether they significantly improve this fit and whether the numerical values of the higher-order bias coefficients that we obtain from the fit are strongly enhanced.

To make these arguments more explicit, we estimate the scaling of the higher-derivative and one-loop terms, taking care to keep track of the bias coefficients that enter with each type of term. For example, consider the one-loop diagram $B_{hhh,321}$ that contributes to the halo-halo-halo bispectrum, where the ``$321$'' subscript indicates that this term is a product of cubic, quadratic, and linear fields, following the notation in \cite{Angulo:2014tfa}. In the high-bias-scaling limit, this contribution scales relative to the tree-level halo-halo-halo bispectrum as
\begin{equation} \frac{B_{hhh,321}}{B_{hhh,\rm tree}} \sim \frac{\langle\delta_{h}^{(3)}\delta_{h}^{(2)}\delta_{h}^{(1)}\rangle}{\langle\delta_{h}^{(2)}\delta_{h}^{(1)}\delta_{h}^{(1)}\rangle} \sim \frac{b_3 b_2 b_1\langle\delta_1^3\delta_2^2\delta_3\rangle}{b_2 b_1^2 \langle\delta_1^2\delta_2\delta_3\rangle} \sim \frac{b_3}{b_1}L, 
\label{b321tree}
\end{equation}
where $L\sim k^3 P(k)$ is the dark-matter loop suppression factor and $\delta_h^{(n)}$ refers to the overdensity of biased tracers evaluated at $n$-th order in perturbation theory. Meanwhile, higher-derivative terms scale relative to the tree-level halo-halo-halo bispectrum as
\begin{equation} \frac{B_{hhh,\rm deriv}}{B_{hhh,\rm tree}} \sim \frac{b_2 b_1^2 \tfrac{k^2}{k_M^2}\langle\delta_1^2\delta_2\delta_3\rangle}{b_2 b_1^2 \langle\delta_1^2\delta_2\delta_3\rangle} \sim \frac{k^2}{k_M^2}. \end{equation}
We can therefore compare the one-loop contribution $B_{hhh,321}$ to the higher-derivative terms directly:
\begin{equation} \frac{B_{hhh,321}}{B_{hhh,\rm deriv}} = \frac{B_{hhh,321}}{B_{hhh,\rm tree}}\frac{B_{hhh,\rm tree}}{B_{hhh,\rm deriv}} \sim \frac{b_3/b_1}{k^2/k_M^2}L. \end{equation}
\cite{Fujita:2016dne} assumes that the ratio $b_3/b_1$ is nearly constant with respect to tracer mass, which implies that $B_{hhh,321}$ becomes a negligible contribution relative to $B_{hhh,\rm deriv}$ in the limit of very massive tracers. However, as we have argued, the assumption that $b_3/b_1$ is a constant or weak function of tracer mass is not \emph{a priori} justified. For example, if $b_3/b_1$ scales as $b_1^2$, then $B_{hhh,321}/B_{hhh,\rm deriv}$ is not guaranteed to be negligible for very massive objects. Thus, we must check whether such potentially bias-enhanced one-loop contributions should be included in the prediction for the bispectra of very massive tracers.

Calculating the full one-loop bispectra for biased tracers, which include terms that are quartic in the fluctuations, is not trivial. However, the calculation simplifies considerably in the high-bias-scaling limit, i.e., assuming that the bias coefficients scale as $b_n \sim b_1^n$. In this limit and for $b_1 \gg 1$, we only need to consider the terms in the one-loop bispectra that enter at the highest order in the bias coefficients. The rest of the one-loop contributions, which are \emph{not} maximally enhanced in this limit, will not grow as quickly with increasing tracer mass, so these terms cannot be responsible for degrading the perturbative reach of the tree-level bispectra under the assumption that $b_n \sim b_1^n$. We consider an explicit example to clarify this point: in the diagram $B_{hhh,321}$, which appeared in \eqn{b321tree}, notice that we can replace $\delta_h^{(3)}$ by $b_1\delta^{(3)}$ or by $b_3(\delta^{(1)})^{3}$ when performing the contraction. The first term will not grow as quickly with tracer mass as $b_3(\delta^{(1)})^{3}$ if $b_3\sim b_1^3$, so it is negligible in the high-bias-scaling limit and we do not need to include it in our calculation. In this way, we can eliminate many of the possible contractions for $B_{hhh,321}$. We will show that similar simplifications occur for the other relevant one-loop contributions.%\footnote{We only calculate the leading parts of the one-loop contributions in the high-bias-scaling limit, and we leave the calculation of the complete one-loop bispectra for biased tracers to future work.}.

We therefore consider the following small, easy-to-compute subset of the full one-loop bispectra for biased tracers. We include the leading terms in $B_{hhh,321}$, since $B_{hhh,321}/B_{hhh,\rm deriv}\propto b_3/b_1$ in the high-bias-scaling limit (\eqn{b321tree}). By a similar argument, we must include the leading contributions from $B_{hhh,222}$, since $B_{hhh,222}/B_{hhh,\rm deriv}\sim (b_2/b_1)^2 \sim b_3/b_1$ in the high-bias-scaling limit. As it is clear, we refer to the contributions that scale as $b_3/b_1 \sim b_1^{n-1}$ as ``maximally enhanced.'' We also include the leading contributions for the one-loop halo-halo-matter diagram $B_{hhm,321}$, since they are also maximally enhanced:
\begin{equation} \frac{B_{hhm,321}}{B_{hhm,\rm tree}} \sim \frac{b_3 b_2}{b_2 b_1}L \sim \frac{b_3}{b_1}L. \end{equation}
Note that $B_{hhm,222}/B_{hhm,\rm{tree}} \sim b_2/b_1$, which is smaller than $b_3/b_1$ in the high-bias-scaling limit; thus, we do not need to include this diagram since it would be subleading. However, we {\it do} include $B_{hmm,321}$ even though $B_{hhm,321}/B_{hhm,\rm{tree}} \sim b_3/b_2$. We include this diagram for several reasons despite the fact that it is subleading with respect to the aforementioned one-loop corrections to the halo-halo-halo and halo-halo-matter bispectra. Most notably, the errors in the simulated halo-matter-matter bispectra that we use to fit our prediction are smaller than the corresponding halo-halo-matter and halo-halo-halo errors by factors of about 4 and 16, respectively. Thus, \emph{if} $b_3$ is much larger than $b_1$ and $b_2$, we expect to obtain a less biased estimate of $b_3$ by including the leading contribution to $B_{hmm}$ since the smaller errorbars associated with the numerical data provide stronger constraints. In addition, including $B_{hmm,321}$ ensures that the cubic bias coefficients appear in all three bispectra that we use, thereby enhancing the constraining power of the fit. Finally, it is not necessary to include $B_{411}$ diagrams for any of the bispectra; in particular, we show in \S\ref{sec:b411} that these diagrams are degenerate with the tree-level bispectra in the high-bias-scaling limit. Thus, including $B_{411}$ would simply renormalize the quadratic biases, which does not affect the $k$-reach of the fit.%the halo-matter-matter one-loop terms scale as $B_{hmm,321}/B_{hmm,tree}\sim b_3/b_2$ and $B_{hmm,222}/B_{hmm,tree}\sim \text{const.}$, and we include both of these terms in our calculation.

\subsection{Leading One-Loop Contributions in the High-Bias-Scaling Limit}
\label{sec:loop}

We now calculate the leading subset of the one-loop contributions to the bispectra of very massive tracers in the high-bias-scaling limit. Note that the ``Basis of Descendants'' (\emph{BoD}) used in \cite{Fujita:2016dne}, which consists of all non-degenerate operators that appear at each perturbative order, is not convenient for our purposes since the \emph{BoD} operators mix biases that are maximally enhanced in the high-bias-scaling limit with non-maximally enhanced biases. For example, consider the cubic \emph{BoD} kernel
\begin{equation} K_{s}^{(3)}(\mathbf{q}_{1},\mathbf{q}_{2},\mathbf{q}_{3},t) \supset \tilde{c}_{s^{2},2}(t)\ \widehat{c}_{s^{2},2}^{(3)}(\mathbf{q}_{1},\mathbf{q}_{2},\mathbf{q}_{3}) \propto c_{s^{2},2}(t) - \frac{3}{4}c_{s^{3}}(t) - \frac{1}{2}c_{st}(t) - \frac{2}{7}c_{\psi}(t), \label{eq:kernel}\end{equation}
where we used the expression given in \cite{Fujita:2016dne} for $\tilde{c}_{s^{2},2}$. Here, $c_{s^{3}}$, $c_{st}$, and $c_{\psi}$ are maximally bias-enhanced coefficients in the high-bias-scaling limit, since they are associated with third-order operators that enter at cubic order in the fluctuations. However, $c_{s^{2},2}$ is \emph{not} maximally bias enhanced since it corresponds to a third-order operator that is only quadratic in the fluctuations, which means that it is less enhanced relative to the other coefficients in \eqn{eq:kernel} in the high-bias-scaling limit. %This also follows from the fact that $c_{s^{2},2}$ is a streaming term. \todo{need to define streaming term}

To calculate the leading one-loop bispectra in the high-bias-scaling limit, we therefore construct a new basis that explicitly includes the full set of maximally bias-enhanced operators at each perturbative order. Up to cubic order in the fluctuations, these maximally-enhanced operators are $\mathbb{C}_{\delta,1}^{(1)}, \mathbb{C}_{\delta^{2},1}^{(2)}, \mathbb{C}_{\delta^{3}}^{(3)}, \mathbb{C}_{s^{3}}^{(3)}, \mathbb{C}_{st}^{(3)}, \mathbb{C}_{\psi}^{(3)},$ and $\mathbb{C}_{\delta s^{2}}^{(3)}$; explicit expressions for these operators are provided in~\cite{Fujita:2016dne}. The bias coefficients corresponding to these operators, which we label $c_{\delta,1},\ c_{\delta^2,1}$, and so on, can be related to the \emph{BoD} bias coefficients, which we denote by $\tilde{c}$'s, by the expressions provided in~\cite{Fujita:2016dne}. In particular, we take the high-bias-scaling limit of the expressions for the \emph{BoD} coefficients in~\cite{Fujita:2016dne} to obtain the following relations:
\begin{eqnarray} \tilde{c}_{\delta,1} &=& c_{\delta,1}, \label{eq:c1}\n
\tilde{c}_{\delta,2} &=& c_{\delta,2} + \tfrac{7}{2}c_{s^{2},1}, \label{eq:c1}\n
\tilde{c}_{\delta,3} &\simeq& \tfrac{45}{4}c_{s^{3}} + \tfrac{9}{2}c_{st} + 2c_{\psi},\label{eq:c2} \n
\tilde{c}_{\delta,3_{c_{s}}} &=& c_{\delta,3_{c_{s}}} + \tfrac{1}{2\pi}\tfrac{1}{c_{s(1)}^{2}}\tfrac{k^2_{\rm{NL}}}{k_M^2}c_{\partial^2 \delta,1},\n
\tilde{c}_{\delta^{2},1} &=& c_{\delta^{2},1} - \tfrac{17}{6}c_{s^{2},1}, \n 
\tilde{c}_{\delta^{2},2} &\simeq& -\tfrac{137}{16}c_{s^{3}} - \tfrac{71}{24}c_{st} - \tfrac{55}{42}c_{\psi} + \tfrac{7}{4}c_{\delta s^{2}}, \n
\tilde{c}_{s^{2},2} &\simeq& -\tfrac{3}{4}c_{s^{3}} - \tfrac{1}{2}c_{st} - \tfrac{2}{7}c_{\psi},\label{eq:c5} \n
\tilde{c}_{\delta^{3}} &=& c_{\delta^{3}} + \tfrac{511}{72}c_{s^{3}} + \tfrac{25}{12}c_{st} + c_{\psi} - \tfrac{17}{6}c_{\delta s^{2}}. \label{eq:c6}\end{eqnarray}
We have suppressed the time dependence of the coefficients in this equation for brevity. The seven maximally-enhanced $c$ coefficients that correspond to the $\mathbb{C}_{\ldots}^{(n)}$ operators listed above and appear up to one-loop order in the high-bias-scaling limit cannot be uniquely expressed in terms of the \emph{BoD} coefficients that appear in the one-loop power spectra and tree-level plus higher-derivative bispectra. This is because including the maximally bias-enhanced contributions to the one-loop bispectra introduces additional bias coefficients relative to the prediction for the one-loop power spectra and tree-level plus higher-derivative bispectra in \cite{Fujita:2016dne}. 
%This is expected, since the leading one-loop contributions to the bispectra in the high-bias-scaling limit happen to be independent of the tree-level bispectra. 
In other words, these contributions break some of the degeneracies used in~\cite{Fujita:2016dne} to construct the \emph{BoD} coefficients. 
Indeed, it is clear from \eqn{eq:c6} that the renormalized bias coefficients that appear in the one-loop power spectra and tree-level plus higher-derivative bispectra (namely ${b}_{\delta,1}$, ${b}_{\delta,2}$, ${b}_{\delta,3}$, ${b}_{{c_{s}}}$, ${b}_{\delta^2}$, and the coefficients corresponding to the stochastic and higher-derivative operators) are identical to our renormalized bias coefficients with the exception of ${b}_{\delta,3}$. This will allow us to reuse the expressions for the one-loop power spectra and tree-level plus higher-derivative bispectra derived in \cite{Fujita:2016dne} up to a simple replacement for $b_{\delta,3}$. We stress again that this replacement is necessary because the additional degeneracy among the \emph{BoD} coefficients used to construct $b_{\delta,3}$ is broken by the inclusion of the leading one-loop contributions in the high-bias-scaling limit. We are left with a total of four additional bias coefficients relative to the Bin 3 fit in \cite{Fujita:2016dne}, and we list these coefficients explicitly in \S\ref{sec:fit}.

We now use our new basis to construct the linear, quadratic, and cubic kernels in the high-bias-scaling limit, following a procedure similar to that in \cite{Fujita:2016dne}. The kernels are given by:
\begin{eqnarray} \tilde{K}_{s}^{(1)}(\mathbf{q}_1,t) &=& c_{\delta,1}(t)\ \widehat{c}_{\delta,1}^{(1)}(\mathbf{q}_1) \label{eq:kernel1}\n
\tilde{K}_{s}^{(2)}(\mathbf{q}_1,\mathbf{q}_2,t) &=& c_{\delta^{2},1}(t)\ \widehat{c}_{\delta^{2},1}^{(2)}(\mathbf{q}_1,\mathbf{q}_2) \label{eq:kernel2}\n
\tilde{K}_{s}^{(3)}(\mathbf{q}_1,\mathbf{q}_2,\mathbf{q}_3,t) &=& c_{\delta^{3}}(t)\ \widehat{c}_{\delta^{3},1}^{(3)}(\mathbf{q}_1,\mathbf{q}_2,\mathbf{q}_3) + c_{s^{3}}(t)\ \widehat{c}_{s^{3},1}^{(3)}(\mathbf{q}_1,\mathbf{q}_2,\mathbf{q}_3) + c_{st}(t)\ \widehat{c}_{st,1}^{(3)}(\mathbf{q}_1,\mathbf{q}_2,\mathbf{q}_3) \n&\phantom{+}& +\ c_{\psi}(t)\ \widehat{c}_{\psi,1}^{(3)}(\mathbf{q}_1,\mathbf{q}_2,\mathbf{q}_3) + c_{\delta s^{2}}(t)\ \widehat{c}_{\delta s^{2},1}^{(3)}(\mathbf{q}_1,\mathbf{q}_2,\mathbf{q}_3). \label{eq:kernel3}
\end{eqnarray}
We refer the reader to \cite{Fujita:2016dne} for the definitions of the $\widehat{c}$ operators. Note that $\widehat{c}_{\delta,1}^{(1)} = \widehat{c}_{\delta^{2},1}^{(2)} = 1$, so the linear and quadratic kernels are simply equal to the coefficients $c_{\delta,1}$ and $c_{\delta^{2},1}$, respectively.

In the following subsections, we provide expressions for the one-loop diagrams $B_{321}$ and $B_{222}$ in the high-bias-scaling limit, and we also show that the leading $B_{411}$ diagrams are degenerate with the tree-level bispectra in this limit. To eliminate the UV-divergences of these one-loop contributions, we take the UV limit $|\mathbf{q}|/|\mathbf{k}| \rightarrow \infty$ of each $q$-integrand derived below by letting $\mathbf{k} \rightarrow \epsilon\mathbf{k}$, taking the limit $\epsilon \rightarrow 0$, and expanding in powers of $\epsilon$. We multiply the coefficients of the various powers of $\epsilon$ in this limit of the integrand by $\Theta(q-k_{\rm UV})$ and we subtract these terms from the respective integrands to obtain the so-called UV-subtracted contributions\footnote{We address IR-divergences in Appendix \ref{sec:IR-safe}.}. Here, $\Theta(x)$ is the Heaviside step function and $k_{\rm UV} = 0.2\ h\ \rm{Mpc}^{-1}$ is the wavenumber above which we subtract the UV contributions. The value of $k_{\rm UV}$ does not affect the result, since a particular choice of $k_{\rm UV}$ simply corresponds to different values for the counterterms. However, this procedure makes the loop integrals and the effects of renormalization numerically smaller. We subtract the $\epsilon^{0}$ and $\epsilon^{2}$ coefficients, which are degenerate with the contributions of the quadratic bias and quadratic higher-derivative bias counterterms for $B_{321}$ and with the contributions of the stochastic and higher-derivative stochastic counterterms for $B_{222}$. The coefficients associated with higher powers of $\epsilon$ would be degenerate with the counterterms implemented by the higher-higher-derivative bias coefficients, which we do not include in our calculation. %Note that the UV contributions to the one-loop bispectra are not negligible, so we expect the linear, stochastic, and higher-derivative bias coefficients to change by order one factors relative to the best-fit values found in \cite{Fujita:2016dne}.

\subsubsection{Calculation of $B_{321}$}

We first consider $B_{321}$. As noted above, the leading diagrams in the high-bias-scaling limit are those that enter at the highest order in the bias coefficients. Thus, we can write cubic halo fields as
\begin{eqnarray} \delta_{h}^{(3)}(\mathbf{k},t) = \iiint \frac{d^{3}q_1}{(2\pi)^{3}}\frac{d^{3}q_2}{(2\pi)^{3}}\frac{d^{3}q_3}{(2\pi)^{3}}\tilde{K}_s^{(3)}(\mathbf{q}_1,\mathbf{q}_2,\mathbf{q}_3,t)\delta_D^{(3)}(\mathbf{k}-\mathbf{q}_1-\mathbf{q}_2-\mathbf{q}_3)\delta({\mathbf{q}_1})\delta({\mathbf{q}_2})\delta({\mathbf{q}_3}),
\end{eqnarray}
with similar expressions for linear and quadratic fields. Proceeding in this way, we obtain the following expression for the leading halo-halo-halo diagram:
\begin{flalign} &B_{hhh,321}(\mathbf{k}_{1},\mathbf{k}_{2},\mathbf{k}_{3},t) = \langle\delta_{h}^{(3)}(\mathbf{k}_{1},t)\delta_{h}^{(2)}(\mathbf{k}_{2},t)\delta_{h}^{(1)}(\mathbf{k}_{3},t)\rangle' + \text{5 permutations} &\n
&= 3!P_{11}(k_{3};t,t)\tilde{K}_{s}^{(1)}(\mathbf{k}_{3},t)\int \frac{d^{3}q}{(2\pi)^{3}}\Big\{\tilde{K}_{s}^{(3)}(-\mathbf{q},-\mathbf{k}_{1}+\mathbf{q},-\mathbf{k}_{3},t)\tilde{K}_{s}^{(2)}(\mathbf{q},\mathbf{k}_{1}-\mathbf{q},t)&\n&\hspace{62.25mm}P_{11}(q;t,t)P_{11}(\lvert \mathbf{k}_{1}-\mathbf{q} \rvert;t,t)\Big\} + \text{5 permutations}.&
\label{eq:b321eq}
\end{flalign}
In the first line, the primed brackets indicate that we are dropping the overall momentum-conserving Dirac $\delta$-function from the expectation value. The $3!$ comes from the $3\times 2\times 1$ possible contractions for a given triangular $(\mathbf{k}_1$, $\mathbf{k}_2, \mathbf{k}_3)$ configuration, while the permutations arise from the different ways to distribute the linear, quadratic, and cubic fields among $\mathbf{k}_1$, $\mathbf{k}_2$, and $\mathbf{k}_3$. Note that we have not included the diagram with an internal contraction among the cubic halo fields in this expression. The diagram that results from contracting two of the $\delta_{h}^{(3)}$ fields into a loop involves an integral of $\tilde{K}_{s}^{(3)}$ over the internal momenta of the loop, and the resulting expression is degenerate with the tree-level bispectrum in the high-bias-scaling limit. Thus, including this diagram would simply renormalize the linear bias coefficients without affecting the perturbative reach of the fit.

\eqn{eq:b321eq} simplifies considerably since the linear and quadratic kernels are constant functions (i.e., $\widehat{c}_{\delta,1}^{(1)} = \widehat{c}_{\delta^{2},1}^{(2)} = 1$). The expressions for the halo-halo-matter and halo-matter-matter diagrams follow immediately from \eqn{eq:b321eq} up to differences in the permutations. In particular, $B_{hhm,321}$ is identical to $B_{hhh,321}$ with the replacement $\tilde{K}_{s}^{(1)}=1$, and $B_{hmm,321}$ is identical to $B_{hhm,321}$ with the replacement $\tilde{K}_{s}^{(2)}=F_{s}^{(2)}$, where $F_{s}^{(2)}$ is the quadratic Eulerian perturbation theory kernel (see~\cite{Bernardeau2002}). Thus, we have:
\begin{flalign} &B_{hmm,321}(\mathbf{k}_{1},\mathbf{k}_{2},\mathbf{k}_{3}) = 3!P_{11}(k_{3})\int \frac{d^{3}q}{(2\pi)^{3}}\big\{\tilde{K}_{s}^{(3)}(-\mathbf{q},-\mathbf{k}_{1}+\mathbf{q},-\mathbf{k}_{3})F_{s}^{(2)}(\mathbf{q},\mathbf{k}_{1}-\mathbf{q})P_{11}(q)P_{11}(\lvert \mathbf{k}_{1}-\mathbf{q} \rvert)\big\}\nonumber&&\n&\hspace{37mm} + (\mathbf{k}_{2}\leftrightarrow \mathbf{k}_{3}\ \text{permutation})\\
&B_{hhm,321}(\mathbf{k}_{1},\mathbf{k}_{2},\mathbf{k}_{3}) = 3!P_{11}(k_{3})c_{\delta^{2},1}\int \frac{d^{3}q}{(2\pi)^{3}}\big\{\tilde{K}_{s}^{(3)}(-\mathbf{q},-\mathbf{k}_{1}+\mathbf{q},-\mathbf{k}_{3})P_{11}(q)P_{11}(\lvert \mathbf{k}_{1}-\mathbf{q} \rvert)\big\}&&\n&\hspace{37mm} + (\mathbf{k}_{1}\leftrightarrow \mathbf{k}_{2}\ \text{permutation})\nonumber&&\\
&B_{hhh,321}(\mathbf{k}_{1},\mathbf{k}_{2},\mathbf{k}_{3}) = 3!P_{11}(k_{3})c_{\delta^{2},1}c_{\delta,1}\int \frac{d^{3}q}{(2\pi)^{3}}\big\{\tilde{K}_{s}^{(3)}(-\mathbf{q},-\mathbf{k}_{1}+\mathbf{q},-\mathbf{k}_{3})P_{11}(q)P_{11}(\lvert \mathbf{k}_{1}-\mathbf{q} \rvert)\big\}&&\n&\hspace{36mm} + \text{5 permutations}.& \label{eq:b321final}
\end{flalign}
Again, we have suppressed the explicit time dependence in this equation. Note that the permutations are fixed by the requirement that halo fields contribute at leading order in the bias coefficients in the high-bias-scaling limit. In addition, matter fields are always evaluated at the lowest perturbative order allowed by the maximally-enhanced bias expansion. We remind the reader that $B_{hmm,321}$ is only next-to-maximally enhanced in the high-bias-scaling limit. We evaluate the integrals in \eqn{eq:b321final} using the expression for $\tilde{K}_{s}^{(3)}$ in \eqn{eq:kernel3}. %We evaluate the integral at each $(\mathbf{k}_1,\mathbf{k}_2,\mathbf{k}_3)$ configuration where we have simulation data for the bispectrum, and we interpolate the results to obtain a piecewise smooth function. \todo{not sure we need this sentence}

\subsubsection{Calculation of $B_{222}$}

Next, consider $B_{222}$. In the high-bias-scaling limit, the leading halo-halo-halo diagram is given by:
\begin{flalign} &B_{hhh,222}(\mathbf{k}_{1},\mathbf{k}_{2},\mathbf{k}_{3},t) = \langle\delta_{h}^{(2)}(\mathbf{k}_{1},t)\delta_{h}^{(2)}(\mathbf{k}_{2},t)\delta_{h}^{(2)}(\mathbf{k}_{3},t)\rangle' &\n
 &= \iiint \frac{d^{3}q}{(2\pi)^{3}}\frac{d^{3}q'}{(2\pi)^{3}}\frac{d^{3}q''}{(2\pi)^{3}}\Big\{\tilde{K}_{s}^{(2)}(\mathbf{q},\mathbf{k}_{1}-\mathbf{q},t)\tilde{K}_{s}^{(2)}(\mathbf{q'},\mathbf{k}_{2}-\mathbf{q'},t)\tilde{K}_{s}^{(2)}(\mathbf{q''},\mathbf{k}_{3}-\mathbf{q''},t)& \n&\hspace{46.7mm}\langle\delta(\mathbf{q})\delta(\mathbf{k}_{1}-\mathbf{q})\delta(\mathbf{q'})\delta(\mathbf{k}_{2}-\mathbf{q'})
\delta(\mathbf{q''})\delta(\mathbf{k}_{3}-\mathbf{q''})\rangle'\Big\}.&\end{flalign}
There are no permutations in this expression because all of the fields are evaluated at quadratic order. There are eight contractions, which lead to identical contributions. Noting that the constant quadratic kernels factor out of the integral and recalling that we only include the halo-halo-halo diagram as we are working in the high-bias-scaling limit, we find:
\begin{flalign} %&B_{hmm,222}(\mathbf{k}_{1},\mathbf{k}_{2},\mathbf{k}_{3}) = 8c_{\delta^{2},1}\int \frac{d^{3}q}{(2\pi)^{3}}\big\{P_{11}(q)P_{11}(\lvert \mathbf{q}-\mathbf{k}_{3} \rvert)P_{11}(\lvert \mathbf{q}+\mathbf{k}_{2} \rvert)\big\}\nonumber&&\n
%&B_{hhm,222}(\mathbf{k}_{1},\mathbf{k}_{2},\mathbf{k}_{3}) = 8c_{\delta^{2},1}^2\int \frac{d^{3}q}{(2\pi)^{3}}\big\{P_{11}(q)P_{11}(\lvert \mathbf{q}-\mathbf{k}_{3} \rvert)P_{11}(\lvert \mathbf{q}+\mathbf{k}_{2} \rvert)\big\}\nonumber&&\n
&B_{hhh,222}(\mathbf{k}_{1},\mathbf{k}_{2},\mathbf{k}_{3}) = 8c_{\delta^{2},1}^3\int \frac{d^{3}q}{(2\pi)^{3}}\big\{P_{11}(q)P_{11}(\lvert \mathbf{q}-\mathbf{k}_{3} \rvert)P_{11}(\lvert \mathbf{q}+\mathbf{k}_{2} \rvert)\big\}. \label{eq:b222final}
\end{flalign}
%\begin{flalign} B_{hhh,222}(\mathbf{k}_{1},\mathbf{k}_{2},\mathbf{k}_{3},t) &= 8c_{\delta^{2},1}^3\int \frac{d^{3}q}{(2\pi)^{3}}\big\{P_{11}(q)P_{11}(\lvert \mathbf{q}-\mathbf{k}_{3} \rvert)P_{11}(\lvert \mathbf{q}+\mathbf{k}_{2} \rvert)\big\}.\label{eq:b222final}\end{flalign}
%Note that the expressions for $B_{hhm,222}$ and $B_{hmm,222}$ are identical to $B_{hhh,222}$, but they contain fewer factors of $c_{\delta^{2},1}$.

\subsubsection{Calculation of $B_{411}$}
\label{sec:b411}

The leading contributions from $B_{411}$ in the high-bias-scaling limit are degenerate with terms in the tree-level bispectra. For example, the leading halo-halo-halo contribution is given by:
\begin{flalign} &B_{hhh,411}(\mathbf{k}_{1},\mathbf{k}_{2},\mathbf{k}_{3},t) = \langle\delta_{h}^{(4)}(\mathbf{k}_{1},t)\delta_{h}^{(1)}(\mathbf{k}_{2},t)\delta_{h}^{(1)}(\mathbf{k}_{3},t)\rangle' + \text{2 permutations} &\n &\sim P_{11}(k_{2};t,t)P_{11}(k_{3};t,t)\tilde{K}_{s}^{(1)}(\mathbf{k}_{2},t)\tilde{K}_{s}^{(1)}(\mathbf{k}_{3},t)\int \frac{d^{3}q}{(2\pi)^{3}}\tilde{K}_{s}^{(4)}(\mathbf{q},\mathbf{k}_{2},\mathbf{k}_{3},-\mathbf{q},t)P_{11}(q;t,t),& \label{eq:b411final}\end{flalign}
where $\tilde{K}_s^{(4)}$ is the maximally bias-enhanced quartic kernel. This expression is degenerate with terms of the form $P(k_{2})P(k_{3})\tilde{K}_{s}^{(1)}(\mathbf{k}_{2})\tilde{K}_{s}^{(1)}(\mathbf{k}_{3})\tilde{K}_{s}^{(2)}(\mathbf{k}_{2},\mathbf{k}_{3})$ in the tree-level halo-halo-halo bispectrum, since $\tilde{K}_{s}^{(4)} \sim \widehat{c}_{\delta^{4},1}^{(4)} \sim \text{const.}$ in the high-bias-scaling limit and $\tilde{K}_{s}^{(2)} \sim \widehat{c}_{\delta^{2},1}^{(2)}$ is also constant in this limit. Note that any \emph{other} operator contained in $\tilde{K}_{s}^{(4)}$ --- for example, $\widehat{c}_{\delta s^3}^{(4)}$ --- would enter in an internal contraction, so we would integrate such an operator over the internal momenta of the loop. The resulting expressions would have the same $(\mathbf{k}_{1},\mathbf{k}_{2},\mathbf{k}_{3})$ dependence as \eqn{eq:b411final} and therefore as the tree-level bispectrum. Nearly identical arguments apply for $B_{hhm,411}$ and $B_{hmm,411}$, which correspond to $B_{hhh,411}$ with some of the $\tilde{K}_{s}^{(1)}$ terms set equal to unity. Thus, we do not need to calculate $B_{411}$, since including this contribution would simply renormalize the quadratic bias coefficients without affecting the perturbative reach of the prediction.

\subsection{Summary of Expressions for the Power Spectra and Bispectra}
\label{sec:theory}

To construct the EFTofLSS prediction for the two- and three-point functions of biased tracers including the leading one-loop contributions to the bispectra in the high-bias-scaling limit, we use the expressions for the one-loop power spectra and tree-level plus higher-derivative bispectra given in~\cite{Fujita:2016dne} along with the expressions for the one-loop bispectra provided in the previous section. The renormalization of the power spectra proceeds as in \cite{Fujita:2016dne}; in particular, since the renormalized bias coefficients that appear in the one-loop power spectra and tree-level plus higher-derivative bispectra are identical to our new coefficients with the exception of $b_{\delta,3}$, we can identify the rest of our renormalized bias coefficients with those in~\cite{Fujita:2016dne}.
 %Note that the renormalization of the power spectra yields the following replacement for $\tilde{c}_{\delta,1}$:
%\begin{eqnarray} \tilde{c}_{\delta,1} \rightarrow b_{\delta,1} - \sigma^{2}(t) \left( -\frac{13}{21}\tilde{c}_{\delta,1} -\frac{34}{21}\tilde{c}_{\delta,2} +\frac{47}{21}\tilde{c}_{\delta,3} -2\tilde{c}_{\delta^{2},1} +\frac{110}{21}\tilde{c}_{\delta^{2},2} +\frac{136}{63}\tilde{c}_{s^{2},2} +3\tilde{c}_{\delta^3} \right).\label{eq:renorm}\end{eqnarray}
%Here, the bias coefficient $b_{\delta,1}$ is intended to be a finite contribution. The renormalization procedure is identical in the high-bias-scaling limit, and the analogous replacement for $c_{\delta,1}$ can be found by taking the high-bias-scaling limit of \eqn{eq:renorm} or by explicitly renormalizing the power spectra using the maximally-enhanced kernels; we have verified that both methods yield the same result. In particular, we find:
%\begin{eqnarray} c_{\delta,1} &\rightarrow& b_{\delta,1} - \sigma^{2}(t) \left(\frac{2}{3}c_{\delta s^{2}} -\frac{16}{63}c_{st} +3c_{\delta^3} \right). \label{eq:renorm2}\end{eqnarray}  
%Since \eqn{eq:renorm2} is simply the high-bias-scaling limit of \eqn{eq:renorm}, we can use the expressions for the power spectra and bispectra provided in \cite{Fujita:2016dne} without changing $b_{\delta,1}$~(\footnote{One might worry that working in a different basis affects the renormalized expression for $b_{\delta,1}$, but it makes no difference whether we renormalize the power spectra and then change basis or vice versa, since the result must be basis independent.}).
The following degeneracy was used to construct $b_{\delta,3}$ following the renormalization of the power spectra in \cite{Fujita:2016dne}:
\begin{eqnarray} \tilde{c}_{\delta,3} + 15\tilde{c}_{s^{2},2} \rightarrow b_{\delta,3}. \label{eq:b3}\end{eqnarray}
This degeneracy is broken by the inclusion of the maximally-enhanced one-loop bispectra in the high-bias-scaling limit, since $\tilde{c}_{\delta,3}$ and $\tilde{c}_{s^{2},2}$ are linear combinations of maximally bias-enhanced coefficients that now need to be treated as independent parameters. We therefore replace $b_{\delta,3}$ with the following expression, which we obtain by using our relations for the maximally-enhanced bias coefficients in the high-bias-scaling limit (\eqn{eq:c2}) in \eqn{eq:b3}:
\begin{eqnarray} b_{\delta,3} \rightarrow -3c_{st} -\frac{16}{7}c_{\psi}. \label{eq:b3new}\end{eqnarray}
Again, $b_{\delta,3}$ is the only bias coefficient appearing in the fit in \cite{Fujita:2016dne} that needs to be re-expressed in terms of our new basis. At this point, all of our bias coefficients are finite contributions, so we relabel them with the letter $b$, following the notation in \cite{Fujita:2016dne}. Our expression for the one-loop halo-matter power spectrum is therefore given by:
\begin{flalign} &P_{hm}(k) = b_{\delta,1}\Big(P_{11}(k) + 2\int \frac{d^{3}q}{(2\pi)^{3}}F_{s}^{(2)}(\mathbf{k}-\mathbf{q},\mathbf{q})\widehat{c}_{\delta,1,s}^{(2)}(\mathbf{k}-\mathbf{q},\mathbf{q})P_{11}(q)P_{11}(\lvert \mathbf{k}-\mathbf{q}\rvert)\nonumber&&\n&\hspace{25.6mm} + 3\ P_{11}(k)\int \frac{d^{3}q}{(2\pi)^{3}}\big(F_{s}^{(3)}(\mathbf{k},-\mathbf{q},\mathbf{q}) + \widehat{c}_{\delta,1,s}^{(3)}(\mathbf{k},-\mathbf{q},\mathbf{q}) + \tfrac{13}{63}\big)P_{11}(q)\Big)\nonumber&&\n&\hspace{16mm} + b_{\delta,2}\ 2\int \frac{d^{3}q}{(2\pi)^{3}}F_{s}^{(2)}(\mathbf{k}-\mathbf{q},\mathbf{q})\big(F_{s}^{(2)}(\mathbf{k}-\mathbf{q},\mathbf{q}) - \widehat{c}_{\delta,1,s}^{(2)}(\mathbf{k}-\mathbf{q},\mathbf{q})\big)P_{11}(q)P_{11}(\lvert \mathbf{k}-\mathbf{q}\rvert)\nonumber&&\n&\hspace{16mm} + (-3b_{st} - \tfrac{16}{7}b_{\psi})\ 3\ P_{11}(k)\int \frac{d^{3}q}{(2\pi)^{3}}\big(\widehat{c}_{\delta,3,s}^{(3)}(\mathbf{k},-\mathbf{q},\mathbf{q}) - \tfrac{47}{63}\big)P_{11}(q)P_{11}(\lvert \mathbf{k}-\mathbf{q}\rvert)\nonumber&&\n&\hspace{16mm} + b_{\delta^2}\ 2\int \frac{d^{3}q}{(2\pi)^{3}}F_{s}^{(2)}(\mathbf{k}-\mathbf{q},\mathbf{q})P_{11}(q)P_{11}(\lvert \mathbf{k}-\mathbf{q}\rvert)\nonumber&&\n&\hspace{16mm} + (b_{c_{s}} + b_{\delta,1})(-2(2\pi))c_{s(1)}^{2}\frac{k^2}{k_{\rm NL}^2}P_{11}(q) + b_{\partial^4 \delta}\frac{k^4}{k_{\rm NL}^4}P_{11}(k).& \label{eq:Phm}
\end{flalign}
Note that we have suppressed the explicit time dependence of the bias coefficients and the linear power spectra that appear in this expression. The one-loop halo-halo power spectrum is given by:
\begin{flalign} &P_{hh}(k) = b_{\delta,1}^{2}\Big(P_{11}(k) + 2\int \frac{d^{3}q}{(2\pi)^{3}}\big[\widehat{c}_{\delta,1,s}^{(2)}(\mathbf{k}-\mathbf{q},\mathbf{q})\big]^{2}P_{11}(q)P_{11}(\lvert \mathbf{k}-\mathbf{q}\rvert)\nonumber&&\n&\hspace{34.3mm} + 6\ P_{11}(k)\int \frac{d^{3}q}{(2\pi)^{3}}\big(\widehat{c}_{\delta,1,s}^{(3)}(\mathbf{k},-\mathbf{q},\mathbf{q}) + \tfrac{13}{63}\big)P_{11}(q)\Big)\nonumber&&\n&\hspace{16mm} + b_{\delta,1}(-3b_{st} - \tfrac{16}{7}b_{\psi})\ 6\ P_{11}(k)\int \frac{d^{3}q}{(2\pi)^{3}}\big(\widehat{c}_{\delta,3,s}^{(3)}(\mathbf{k},-\mathbf{q},\mathbf{q}) - \tfrac{47}{63}\big)P_{11}(q)\nonumber&&\n&\hspace{16mm} + b_{\delta,2}^{2}\ 2\int \frac{d^{3}q}{(2\pi)^{3}}\big[F_{s}^{(2)}(\mathbf{k}-\mathbf{q},\mathbf{q}) - \widehat{c}_{\delta,1,s}^{(2)}(\mathbf{k}-\mathbf{q},\mathbf{q})\big]^{2}P_{11}(q)P_{11}(\lvert \mathbf{k}-\mathbf{q}\rvert)\nonumber&&\n&\hspace{16mm} + b_{\delta^2}^{2}\ 2\int \frac{d^{3}q}{(2\pi)^{3}}P_{11}(q)P_{11}(\lvert \mathbf{k}-\mathbf{q}\rvert)\nonumber&&\n&\hspace{16mm} + b_{\delta,1}b_{\delta,2}\ 4\int \frac{d^{3}q}{(2\pi)^{3}}\widehat{c}_{\delta,1,s}^{(2)}(\mathbf{k}-\mathbf{q},\mathbf{q})(F_{s}^{(2)}(\mathbf{k}-\mathbf{q},\mathbf{q}) - \widehat{c}_{\delta,1,s}^{(2)}(\mathbf{k}-\mathbf{q},\mathbf{q}))P_{11}(q)P_{11}(\lvert \mathbf{k}-\mathbf{q}\rvert)\nonumber&&\n&\hspace{16mm} + b_{\delta,1}b_{\delta^2}\ 4\int \frac{d^{3}q}{(2\pi)^{3}}\widehat{c}_{\delta,1,s}^{(2)}(\mathbf{k}-\mathbf{q},\mathbf{q})P_{11}(q)P_{11}(\lvert \mathbf{k}-\mathbf{q}\rvert)\nonumber&&\n&\hspace{16mm} + b_{\delta,2}b_{\delta^2}\ 4\int \frac{d^{3}q}{(2\pi)^{3}}(F_{s}^{(2)}(\mathbf{k}-\mathbf{q},\mathbf{q}) - \widehat{c}_{\delta,1,s}^{(2)}(\mathbf{k}-\mathbf{q},\mathbf{q}))P_{11}(q)P_{11}(\lvert \mathbf{k}-\mathbf{q}\rvert)\nonumber&&\n&\hspace{16mm} + b_{\delta,1}b_{c_{s}}2(-2(2\pi))c_{s(1)}^{2}\frac{k^2}{k_{\rm NL}^2}P_{11}(q) + 2b_{\delta,1}b_{\partial^4 \delta}\frac{k^4}{k_{\rm NL}^4}P_{11}(k)\nonumber&&\n&\hspace{16mm} + b_{\epsilon} - 2b_{\epsilon}b_{\partial^2_{\epsilon}}\frac{k^2}{k_M^4}\nonumber&&\n&\hspace{16mm} - 2\Sigma^2(b_{\delta,1}^2 + b_{\delta,2}^2 - 2b_{\delta,1}b_{\delta,2} - 2b_{\delta,1}b_{\delta^2} + 2b_{\delta,2}b_{\delta^2}),& \label{eq:Phh}
\end{flalign}
where we used
\begin{equation}\Sigma^2(t) = \int\frac{d^{3}q}{(2\pi)^{3}}[P_{11}(q;t,t)]^2.\end{equation}
Expressions for the $\widehat{c}$ operators are provided in~\cite{Fujita:2016dne}.

Our expressions for the tree-level plus higher-derivative plus one-loop bispectra in the high-bias-scaling limit are given by:
\begin{flalign} &B_{hmm}(\mathbf{k}_1,\mathbf{k}_2,\mathbf{k}_3) = 2K_{s}^{(2)}(\mathbf{k}_2,\mathbf{k}_3)P_{11}(k_2)P_{11}(k_3)\nonumber&&\n&\hspace{32.5mm} + 2K_{s}^{(1)}(\mathbf{k}_1)[F_{s}^{(2)}(\mathbf{k}_1,\mathbf{k}_2)P_{11}(k_1)P_{11}(k_2) + F_{s}^{(2)}(\mathbf{k}_1,\mathbf{3}_2)P_{11}(k_1)P_{11}(k_3)] \nonumber&&\n&\hspace{32.5mm} + B_{hmm,321}(\mathbf{k}_1,\mathbf{k}_2,\mathbf{k}_3), %+ B_{hmm,222}(\mathbf{k}_1,\mathbf{k}_2,\mathbf{k}_3)
\end{flalign}
\begin{flalign} &B_{hhm}(\mathbf{k}_1,\mathbf{k}_2,\mathbf{k}_3) = 2K_{s}^{(2)}(\mathbf{k}_2,\mathbf{k}_3)K_{s}^{(1)}(\mathbf{k}_2)P_{11}(k_2)P_{11}(k_3)\nonumber&&\n&\hspace{31.8mm} + 2K_{s}^{(2)}(\mathbf{k}_1,\mathbf{k}_3)K_{s}^{(1)}(\mathbf{k}_1)P_{11}(k_1)P_{11}(k_3)\nonumber&&\n&\hspace{31.8mm} + 2F_{s}^{(2)}(\mathbf{k}_1,\mathbf{k}_2)K_{s}^{(1)}(\mathbf{k}_1)K_{s}^{(1)}(\mathbf{k}_2)P_{11}(k_1)P_{11}(k_2)\nonumber&&\n&\hspace{31.8mm} + b_{\epsilon}P_{11}(k_3)\big(2b_{\epsilon\delta} - b_{\partial^2\epsilon\delta}\tfrac{k_1^2 + k_2^2}{k_M^2} - 2b_{\epsilon\partial^2\delta} \big) \nonumber&&\n&\hspace{31.8mm} + B_{hhm,321}(\mathbf{k}_1,\mathbf{k}_2,\mathbf{k}_3), %+ B_{hmm,222}(\mathbf{k}_1,\mathbf{k}_2,\mathbf{k}_3)
\end{flalign}
\begin{flalign} &B_{hhh}(\mathbf{k}_1,\mathbf{k}_2,\mathbf{k}_3) = \{2K_{s}^{(2)}(\mathbf{k}_1,\mathbf{k}_2)K_{s}^{(1)}(\mathbf{k}_1)K_{s}^{(1)}(\mathbf{k}_1)P_{11}(k_1)P_{11}(k_2) \nonumber&&\n&\hspace{30.8mm} + b_{\epsilon}P_{11}(k_1)\big(2b_{\delta,1}b_{\epsilon\delta} - 2\tfrac{k_1^2}{k_M^2}b_{\partial^2\delta}b_{\epsilon\delta} - \tfrac{k_2^2 + k_3^2}{k_M^2}b_{\delta,1}b_{\partial^2\epsilon\delta} - 2\tfrac{k_1^2}{k_M^2}b_{\delta,1}b_{\epsilon\partial^2\delta}\big) \nonumber&&\n&\hspace{30.8mm} + \text{2 permutations}\} + b_{\epsilon}^2 \nonumber&&\n&\hspace{30.8mm} + B_{hhh,321}(\mathbf{k}_1,\mathbf{k}_2,\mathbf{k}_3) + B_{hhh,222}(\mathbf{k}_1,\mathbf{k}_2,\mathbf{k}_3). %+ B_{hmm,222}(\mathbf{k}_1,\mathbf{k}_2,\mathbf{k}_3)
\end{flalign}
Explicit expressions for the one-loop contributions in the high-bias-scaling limit are provided in \eqn{eq:b321final} and \eqn{eq:b222final}. Since we already subtracted the UV contributions from the bispectra, we can identify the bias coefficients entering the bispectra directly with the renormalized ones. Thus, for the coefficients that enter the bispectra through halo kernels such as $K_s^{(1)}$, we take $c_{\delta,1}\rightarrow b_{\delta,1}$ and we perform corresponding replacements for the rest of the coefficients. We refer the reader to the expressions for the $\emph{BoD}$ kernels in \cite{Fujita:2016dne}. Note that the \emph{BoD} coefficients that appear in $K_s^{(1)}$ and $K_s^{(2)}$ are identical to the coefficients in our new basis, so there is no ambiguity when performing these replacements.
%\begin{eqnarray} B_{hhh}(\mathbf{k}_1,\mathbf{k}_2,\mathbf{k}_3,t) = B_{hhh,tree}(\mathbf{k}_1,\mathbf{k}_2,\mathbf{k}_3,t) + B_{hhh,1-loop}(\mathbf{k}_1,\mathbf{k}_2,\mathbf{k}_3,t).\end{eqnarray}
%Here, $B_{hhh,tree}$ is identical to the expression for the halo-halo-halo bispectrum in \cite{Fujita:2016dne} and $B_{hhh,1-loop}$ is given by
%\begin{eqnarray} B_{hhh,1-loop}(\mathbf{k}_1,\mathbf{k}_2,\mathbf{k}_3,t) = B_{hhh,321}(\mathbf{k}_1,\mathbf{k}_2,\mathbf{k}_3,t) + B_{hhh,222}(\mathbf{k}_1,\mathbf{k}_2,\mathbf{k}_3,t),\label{eq:Bhhhloop}\end{eqnarray}
%where $B_{hhh,321}$ and $B_{hhh,222}$ are given by \eqn{eq:b321eq} and \eqn{eq:b222eq}, respectively.

%----------------------------------------------------------------------------------------
%	SECTION 3
%----------------------------------------------------------------------------------------

\section{Comparison to Simulation Data}

\subsection{Fitting Procedure}
\label{sec:fit}

We now fit the two- and three-point functions from the Millennium-XXL N-body simulation using our prediction for the one-loop power spectra and tree-level plus higher-derivative plus one-loop bispectra of biased tracers in the high-bias-scaling limit. Our fitting procedure, which utilizes the Mathematica routine \emph{NonlinearModelFit}, is very similar to the one described in \cite{Angulo:2015eqa} and summarized in \cite{Fujita:2016dne}, so we refer the reader to those papers for details. To include the leading one-loop contributions to the bispectra in the high-bias-scaling limit, we fit the two- and three-point functions using the following eleven bias coefficients for both light tracers and very massive tracers: $b_{\delta,1}, b_{\delta,2}, b_{\delta^{2}}, b_{c_{s}}, b_{\epsilon\delta}, b_{\epsilon}, b_{\delta^{3}}, b_{s^{3}}, b_{st}, b_{\psi}, b_{\delta s^{2}}$. When including higher-derivative terms for very massive tracers, we fit for the same partial set of higher-derivative bias coefficients used for the Bin 3 fit in~\cite{Fujita:2016dne}: ${b_{\partial^2\delta^2}, b_{(\partial\delta)^2}, b_{\partial^2\epsilon}, b_{\partial^2\epsilon\delta}, b_{\epsilon\partial^2\delta}}$. Choosing the same set of higher-derivative bias coefficients allows us to compare our results directly to those in \cite{Fujita:2016dne}, and therefore to determine whether the subset of one-loop contributions derived above should be included alongside higher-derivative biases in the EFTofLSS prediction for the bispectra of very massive tracers.

Our prediction for the two- and three-point functions of very massive tracers, which includes both higher-derivative biases and the leading one-loop contributions to the bispectra in the high-bias-scaling limit, has four additional bias coefficients relative to the Bin 3 fit in \cite{Fujita:2016dne}. This motivates us to perform the fit for light tracers (Bin 0) using the same subset of one-loop contributions to the bispectra. While we do not expect these one-loop contributions to significantly improve the $k$-reach of the prediction for light tracers, redoing the fit is useful because it gives us a sense of whether our Bin 3 fit improves because we have specifically included the leading one-loop contributions in the high-bias-scaling limit or simply because we have increased the functional freedom of the prediction. In general, we expect the $k$-reach of our Bin 3 prediction to increase relative to the result in \cite{Fujita:2016dne} because we include a subset of the one-loop bispectra, but if there is a comparable improvement in our Bin 0 fit then we would conclude that the additional freedom provided by these diagrams, rather than the bias enhancement of the contributions derived above, improves the prediction for very massive tracers. However, if the $k$-reach of our Bin 3 fit improves significantly with respect to the result in \cite{Fujita:2016dne} and the $k$-reach of our Bin 0 prediction is roughly the same as the $k$-reach of the tree-level Bin 0 bispectra, and at the same time the values of the bias coefficients that we obtain are consistent with the enhanced scaling $b_n\sim b_1^n$, this would indicate that the one-loop contributions derived above are strongly bias enhanced and that these contributions must be included in order to accurately fit the bispectra for very massive tracers above $k\simeq 0.12\ h\ \rm{Mpc}^{-1}$.

\subsection{Results} 
\label{sec:results}

\begin{figure}[htb]
\centering
\includegraphics[width=1.0\columnwidth]{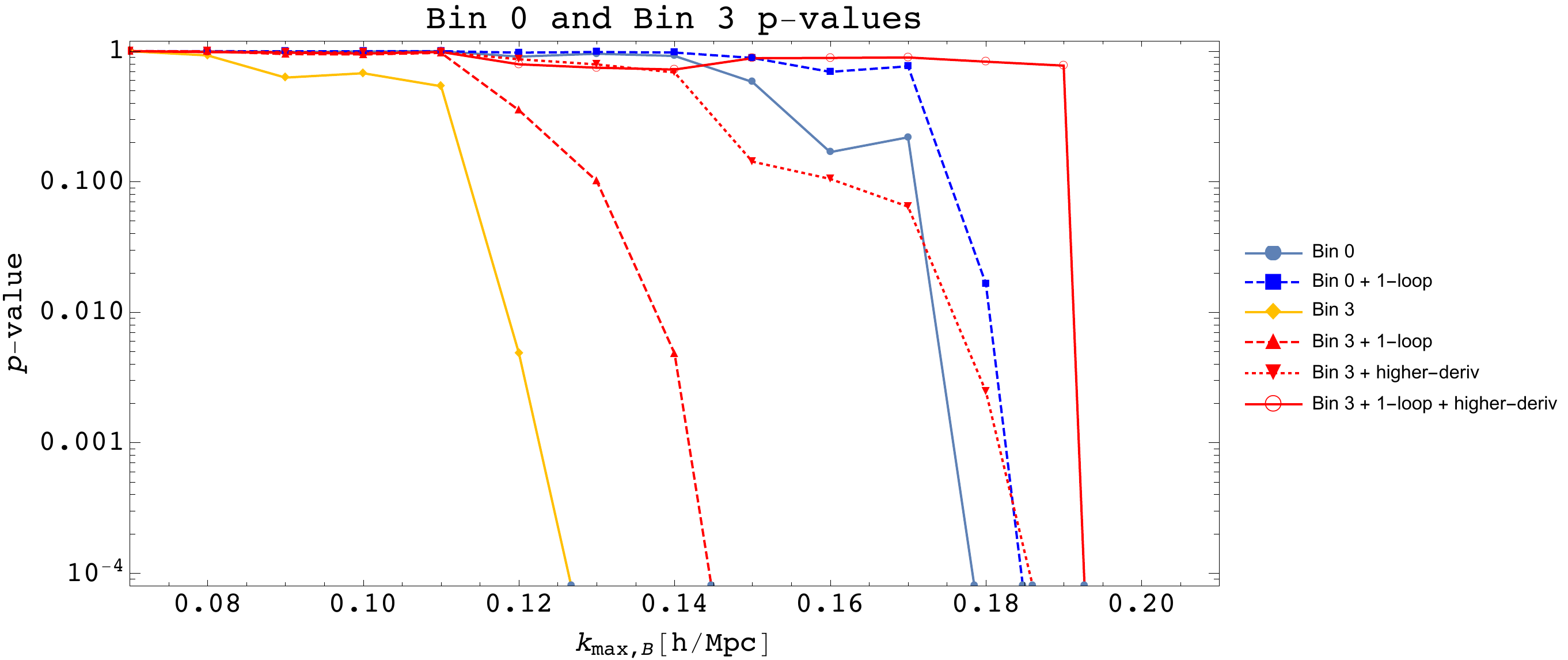} \caption{Results from fitting EFTofLSS predictions for the power spectra and bispectra of biased tracers to data from the Millennium-XXL N-body simulation. We plot $p$-values from fits to the two- and three-point functions of light tracers (Bin 0; $\rm{M}_{\rm halo} \sim 10^{12}\ \rm{M}_{\rm \odot}$) and very massive tracers (Bin 3; $\rm{M}_{\rm halo} \sim 10^{14}\ \rm{M}_{\rm \odot}$). For Bin 0, we show $p$-values for the with the one-loop power spectra and tree-level bispectra (light blue; $k_{\rm{fit}} = 0.15\ h\ \rm{Mpc}^{-1}$), and for the fit where the bisepctra include the leading one-loop contributions in the high-bias-scaling limit (dashed blue; $k_{\rm{fit}} = 0.15\ h\ \rm{Mpc}^{-1}$). For Bin 3, we show $p$-values for the fit with the one-loop power spectra and tree-level bispectra (yellow; $k_{\rm{fit}} = 0.12\ h\ \rm{Mpc}^{-1}$), and for fits that include the leading one-loop bispectra in the high-bias-scaling limit (dashed red; $k_{\rm{fit}} = 0.14\ h\ \rm{Mpc}^{-1}$), higher-derivative contributions to the bispectra (dotted red; $k_{\rm{fit}} = 0.12\ h\ \rm{Mpc}^{-1}$), and both higher-derivative biases and the leading one-loop contributions (solid red; $k_{\rm{fit}} = 0.15\ h\ \rm{Mpc}^{-1}$). We argue based on these fits that higher-derivative biases are the leading correction to the tree-level bispectra for very massive tracers.}%As described in \cite{Fujita:2016dne}, $p$-values are obtained using the best-fit bias coefficients evaluated at $k_{\rm{fit}}$.}
\label{fig:pvalue}
\end{figure}

We now present our results. The $p$-values for our Bin 3 fit that includes both higher-derivative terms and the leading one-loop contributions to the bispectra in the high-bias-scaling limit are shown by the solid red line in Figure \ref{fig:pvalue}. We also plot $p$-values for several other Bin 3 calculations: the fit with the one-loop power spectra and tree-level bispectra corresponds to the yellow line, the fit that includes only higher-derivative biases corresponds to the dotted red line, and the fit that includes the one-loop contributions to the bispectra derived above but does \emph{not} include higher-derivative terms corresponds to the dashed red line. The $p$-values for our Bin 0 fit that includes the leading one-loop contributions to the bispectra in the high-bias-scaling limit are shown by the dashed blue line, and the $p$-values for the Bin 0 fit with the bispectra evaluated at tree level are shown by the solid blue line. We calculate $p$-values by using the best-fit bias coefficients evaluated at $k_{\text{fit}}$, where $k_{\text{fit}}$ is the smallest wavenumber at which the best-fit value for any of the bias coefficients is discrepant from the best-fit value at any lower wavenumber by more than one standard deviation. We plot $p$-values as a function of $k_{\text{max},B}$, which is the maximum wavenumber among all of the datapoints used to determine the best-fit bias coefficients as functions of $k$. We refer the reader to \cite{Fujita:2016dne} for a detailed description of the method for calculating $p$-values.

\begin{table}[htb]
\begin{center}
\begin{tabular}{ |P{3cm}||P{3cm}|P{3cm}| }
\hline
\multicolumn{3}{|c|}{Bias Coefficient Values and Errors} \\
\hline
Parameter &Bin 0 &Bin 3\\
\hline
$b_{\delta,1}$   &$1.00 \pm 0.06$& $2.91 \pm 0.02$\\
$b_{\delta,2}$&$-0.05 \pm 0.08$& $1.47 \pm 0.91$\\
$b_{\delta,3}$ &$0.62 \pm 0.16$& $2.42 \pm 0.90$\\
$b_{\delta^2}$    &$0.57 \pm 0.08$& $1.62 \pm 0.76$\\
$b_{c_s}$&$1.23 \pm 0.33$& $-7.47 \pm 1.47$\\
$b_{\epsilon\delta}$&$0.49 \pm 0.11$& $1.62 \pm 0.14$\\
$b_\epsilon$&$5780 \pm 147$& $121000 \pm 1360$\\
$b_{\delta^3}$&$-0.35 \pm 0.08$& $-1.59 \pm 0.55$\\
$b_{s^3}$&$-0.47 \pm 0.71$& $11.9 \pm 8.38$\\
$b_{st}$&$-0.45 \pm 0.12$& $-2.56 \pm 0.65$\\
$b_{\psi}$&$0.32 \pm 0.11$& $2.30 \pm 0.62$\\
$b_{\delta s^2}$&$0.52 \pm 0.12$& $2.05 \pm 0.91$\\
$b_{\partial^2\delta^{2}}/k_{M}^2$& --  & $103 \pm 24.3$\\
$b_{(\partial\delta)^{2}}/k_{M}^2$& --  & $-147 \pm 42.2$\\
$b_{\partial^2\epsilon}/k_{M}^2$& --  & $0.07 \pm 0.06$\\
$b_{\partial^2\epsilon\delta}/k_{M}^2$& --  & $-5.78 \pm 8.20$\\
$b_{\epsilon\partial^2\delta}/k_{M}^2$& --  & $50.8 \pm 15.9$\\
\hline
\end{tabular}
\caption{Best-fit bias coefficients for our Bin 0 and Bin 3 fits, measured in units of $h\ \text{Mpc}^{-1}$ to the appropriate power. Both fits include the leading one-loop contributions to the bispectra in the high-bias-scaling limit, which enter with the bias coefficients $b_{\delta^3}$, $b_{s^3}$, $b_{st}$, $b_{\psi}$, and $b_{\delta s^2}$, and the Bin 3 fit also includes the same partial set of higher-derivative terms used in \cite{Fujita:2016dne}. We list the inferred value of $b_{\delta,3} = -3b_{st} - 16b_{\psi}/7$ for comparison with the results in \cite{Fujita:2016dne}, but we note that $b_{\delta,3}$ is \emph{not} an independent parameter in these fits. We do not find strong evidence for the high-bias-scaling relation $b_n/b_1 \gg 1$, which implies that higher-derivative terms are the leading correction to the tree-level bispectra for very massive tracers.}%As in \cite{Fujita:2016dne}, higher-derivative terms are not required to fit the Bin 0 simulation data up to the $k_{\text{max}}$ of interest.}
\label{tab:values}
\end{center}
\end{table}

Figure \ref{fig:pvalue} shows that the leading one-loop contributions to the bispectra in the high-bias-scaling limit mildly improve the prediction for very massive tracers: we find a $k$-reach of $\sim 0.19\ h\ \rm{Mpc}^{-1}$ for the fit with both higher-derivative biases and these one-loop contributions, which is somewhat higher than the $k$-reach of $\sim 0.17\ h\ \rm{Mpc}^{-1}$ for the fit with the tree-level plus higher-derivative bispectra\footnote{Note that we do \emph{not} arbitrarily increase the value of $k_{\text{fit}}$ for any of our results, in contrast to \cite{Fujita:2016dne}.}. Similarly, the Bin 3 fit in which the bispectra are evaluated at tree level fits the data up to $\sim 0.12\ h\ \rm{Mpc}^{-1}$, which improves to $\sim 0.14\ h\ \rm{Mpc}^{-1}$ when the leading one-loop contributions in the high-bias-scaling limit are included. The Bin 0 results in Figure \ref{fig:pvalue} show that we obtain a similar improvement for light tracers, indicating that these improvements are mainly due to the increased functional freedom associated with the one-loop contributions rather than a strong bias enhancement. In particular, the $k$-reach of the Bin 0 fit with the tree-level bispectra ($\sim 0.17\ h\ \rm{Mpc}^{-1}$) improves by about $0.01\ h\ \rm{Mpc}^{-1}$ when the leading one-loop contributions in the high-bias-scaling limit are included, which is roughly the same amount of improvement we find for Bin 3. Thus, the leading one-loop contributions to the bispectra in the high-bias-scaling limit improve the prediction comparably for light tracers and very massive tracers, which implies that the high-mass tracers we consider here do not obey the high-bias-scaling relation $b_n/b_1 \gg 1$.

This interpretation is consistent with and further supported by the fact that the best-fit values for the quadratic and cubic bias coefficients are generally comparable to the best-fit values for the linear biases in our Bin 3 fit that includes both higher-derivative biases and the leading one-loop contributions to the bispectra in the high-bias-scaling limit, with the possible exception of $b_{s^3}$, which however is affected by larger uncertainties. Thus, we do not find strong evidence that $b_n \sim b_1^{n}$ for these tracers\footnote{Moreover, we do not find strong evidence that this scaling holds when only higher-derivative or one-loop contributions are included.}. This can be seen from Table \ref{tab:values}, which lists best-fit values and error estimates for the bias coefficients corresponding to the Bin 0 and Bin 3 fits that include both higher-derivative biases (for Bin 3) and the leading one-loop contributions to the bispectra. The bias coefficients in Table \ref{tab:values} are evaluated at $k_{\text{fit}} = 0.15\ h\ \rm{Mpc}^{-1}$ for both of these fits. The correlation matrices corresponding to the bias coefficients in these fits are shown in Tables \ref{tab:Bin0correlation} and \ref{tab:Bin3correlation}, respectively. %From Table \ref{tab:Bin3correlation}, we see that the one-loop bias coefficients in the Bin 0 fit are nearly completely correlated or anti-correlated, which confirms that these terms are unnecessary for light tracers. 
%We note that the errors for some of the bias coefficients introduced by the one-loop contributions are comparable to the best-fit values for these coefficients, which implies that certain coefficients are not well constrained by the simulated bispectra, which only extend to $k = 0.14\ h\ \rm{Mpc}^{-1}$. %We test this by redoing the fits with these coefficients set to zero and checking that the $p$-values are approximately unchanged relative to the fits that include all of the leading one-loop contributions in the high-bias-scaling limit. As expected, we find that the resulting $p$-values are nearly identical up to $k \sim 0.17\ h\ \rm{Mpc}^{-1}$.

Although Figure \ref{fig:pvalue} shows that higher-derivative terms provide the leading correction to the tree-level bispectra for very massive tracers, the leading one-loop contributions in this limit are not extremely suppressed (though they do not obey the high-bias-scaling relation). Thus, both the $k$-reach and the overall goodness of fit for the calculation that includes both higher-derivative terms and the leading one-loop contributions in the high-bias-scaling limit improve relative to the fit with the tree-level plus higher-derivative bispectra.

In Figure \ref{fig:Bin3traces}, we plot the best-fit bias coefficients for the Bin 3 fit that includes both higher-derivative terms and the leading one-loop contributions to the bispectra in the high-bias-scaling limit. Similarly, Figure \ref{fig:Bin0traces} shows the best-fit bias coefficients for the Bin 0 fit that includes these one-loop contributions. The bias coefficients for the Bin 3 fit have converged for $k_{\text{max},B}\gtrsim 0.12\ h\ \rm{Mpc}^{-1}$; however, as in \cite{Fujita:2016dne}, we find unexpectedly large errors for smaller values of $k_{\text{max},B}$. Clearly, including the one-loop contributions derived above does not eliminate this issue.%Most of the bias coefficients associated with the one-loop contributions in the Bin 0 and Bin 3 fits converge to similar values within the relevant uncertainties, which again implies that massive tracers are not strongly bias enhanced.

\begin{figure}[htb]
\centering
\includegraphics[width=1.0\columnwidth]{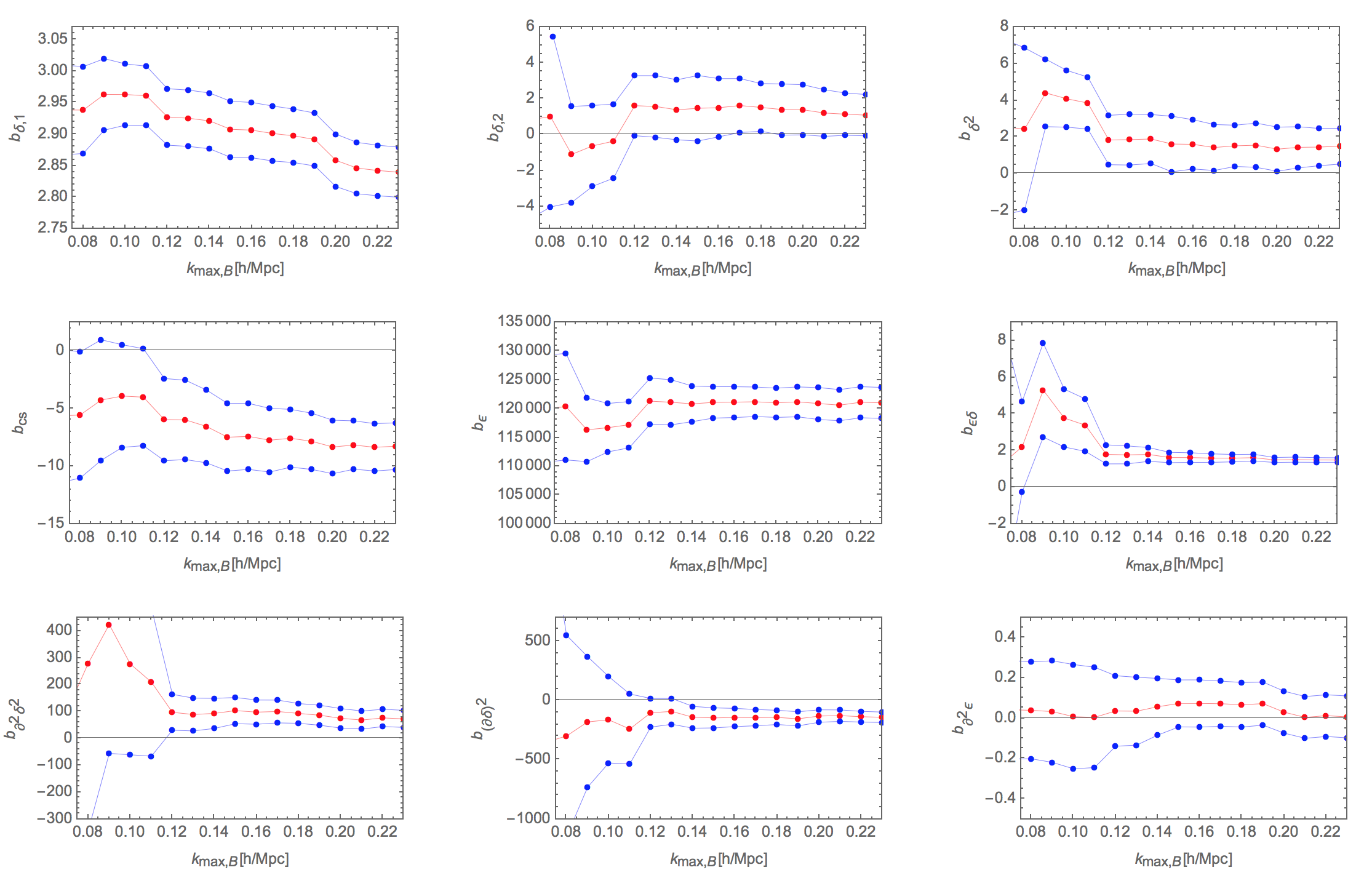}
\includegraphics[width=1.0\columnwidth]{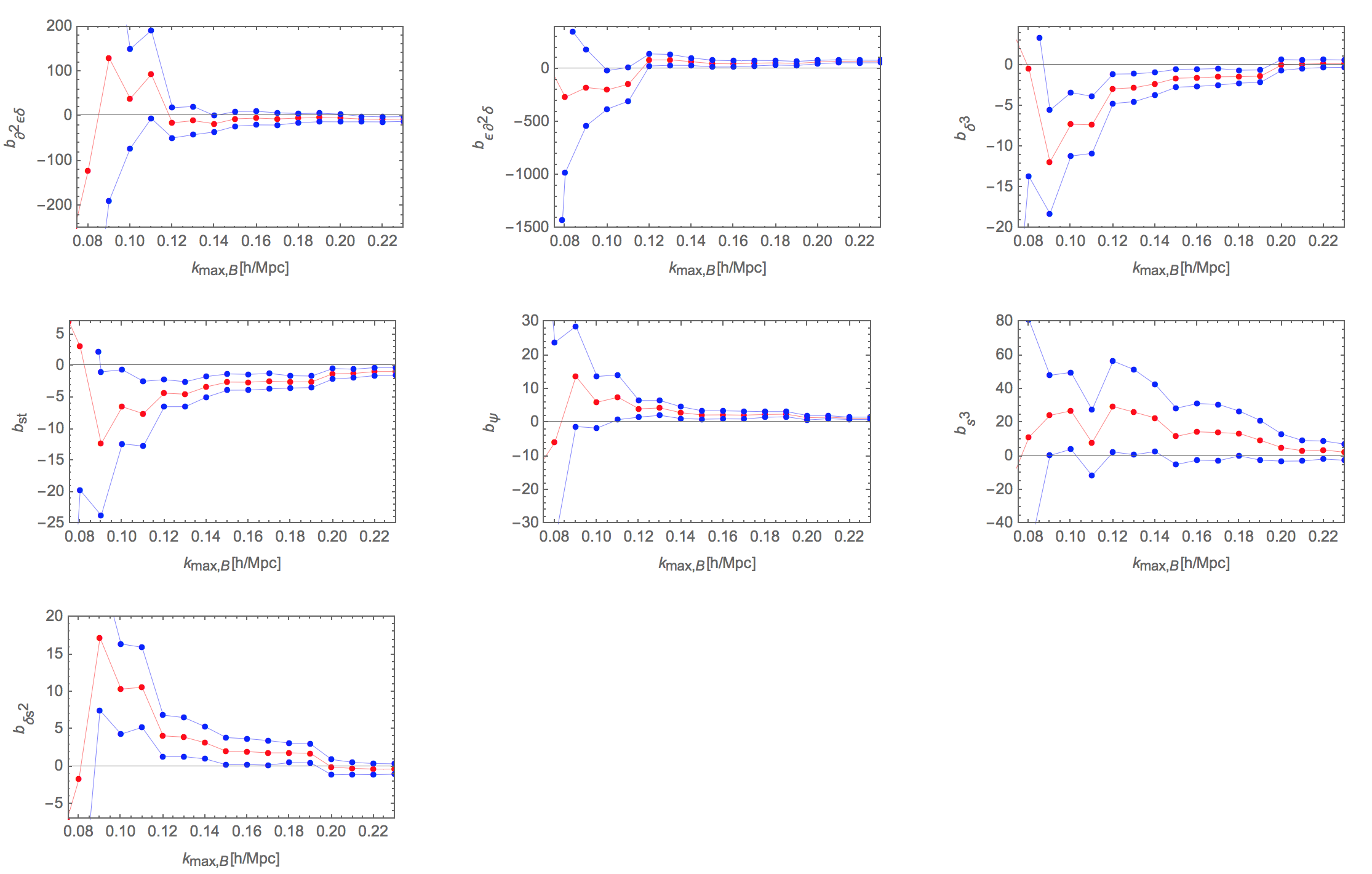}
\caption{Values of the bias coefficients for our Bin 3 fit with the one-loop power spectra and tree-level plus higher-derivative plus one-loop bispectra in the high-bias-scaling limit as functions of $k_{\text{max},B}$. As in \cite{Fujita:2016dne}, there is some unexpected behavior for $k_{\text{max},B}\lesssim 0.12\ h\ \rm{Mpc}^{-1}$ that we ignore to fix the best-fit values of the bias coefficients. Following the procedure outlined in the text, we find $k_{\text{fit}} = 0.15\ h\ \rm{Mpc}^{-1}$ for this fit.}
\label{fig:Bin3traces}
\end{figure}

\begin{figure}[htb]
\centering
\includegraphics[width=1.0\columnwidth]{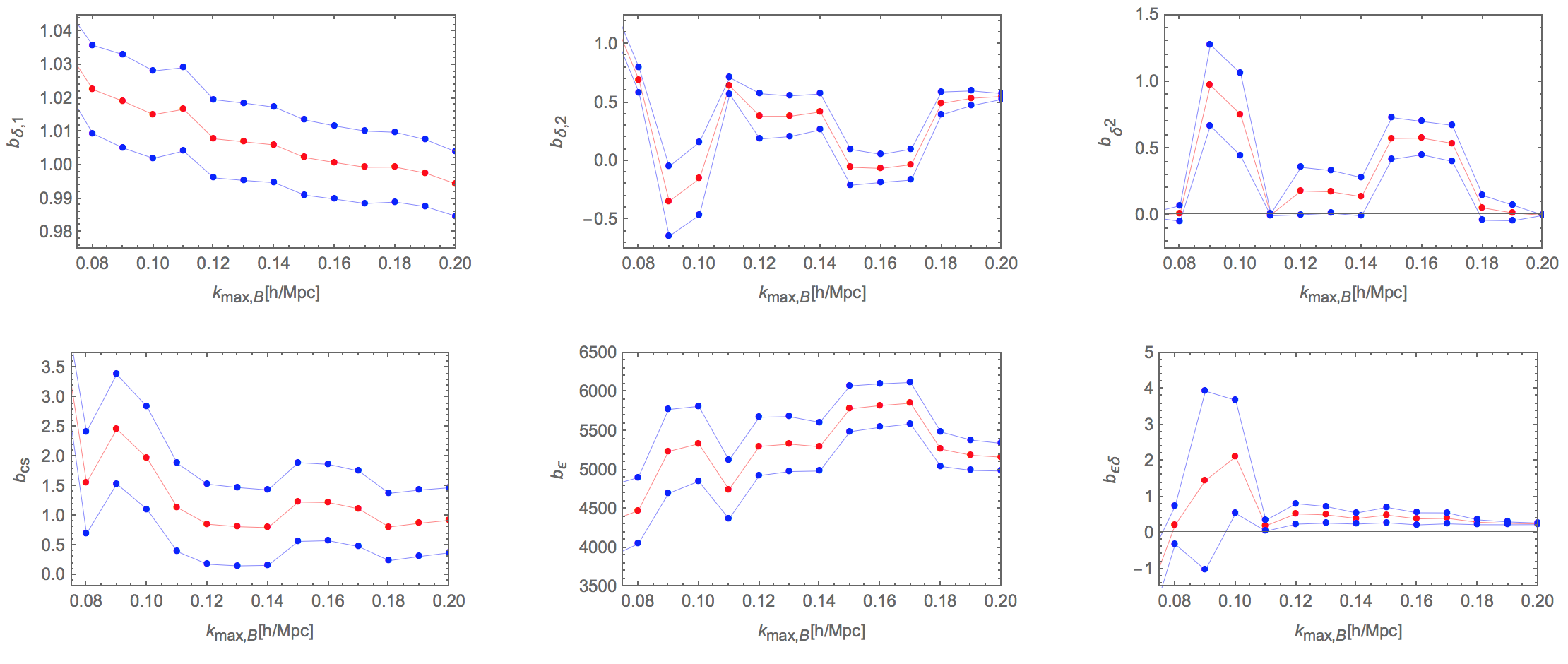}
\includegraphics[width=1.0\columnwidth]{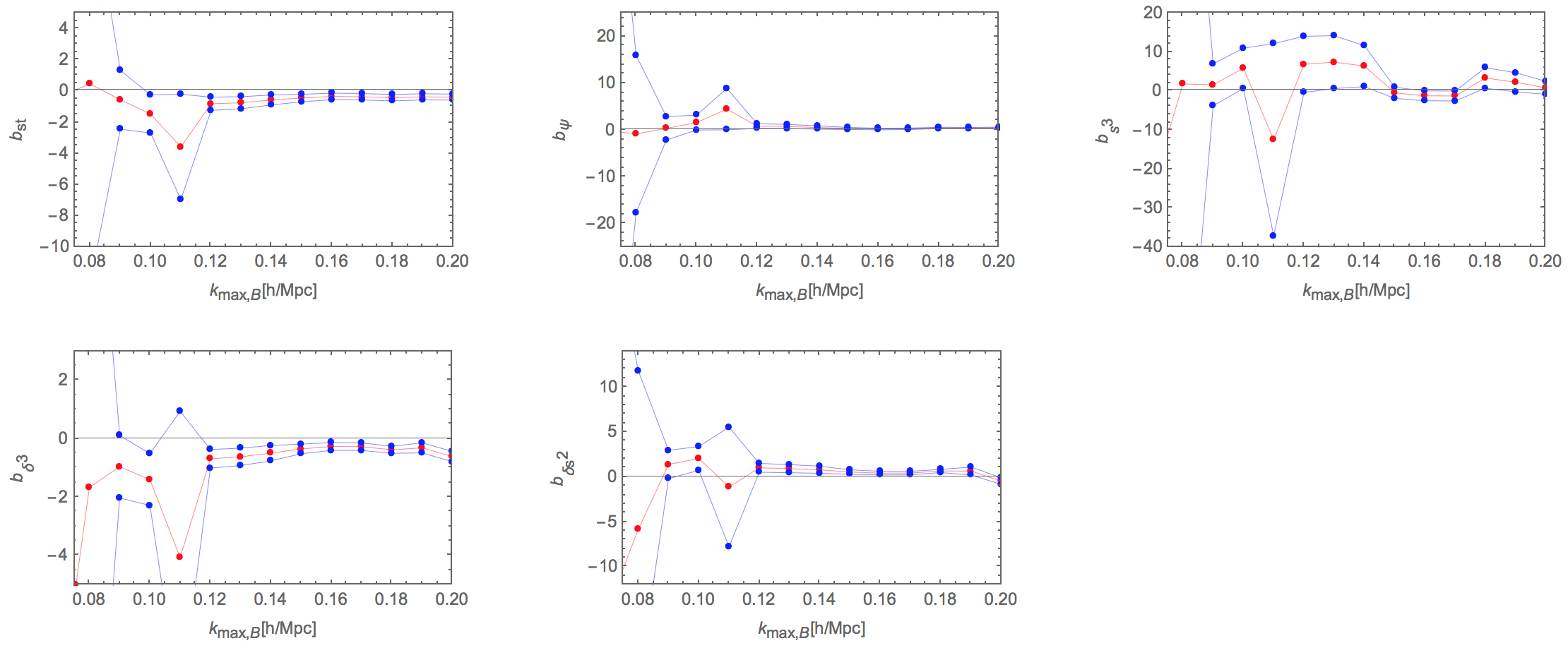}
\caption{Values of the bias coefficients for our Bin 0 fit with the one-loop power spectra and tree-level plus one-loop bispectra in the high-bias-scaling limit as functions of $k_{\text{max},B}$. Following the procedure outlined in the text, we find $k_{\text{fit}} = 0.15\ h\ \rm{Mpc}^{-1}$ for this fit.}
\label{fig:Bin0traces}
\end{figure}

\begin{figure}[htb!]
\centering
\includegraphics[width=1.0\columnwidth]{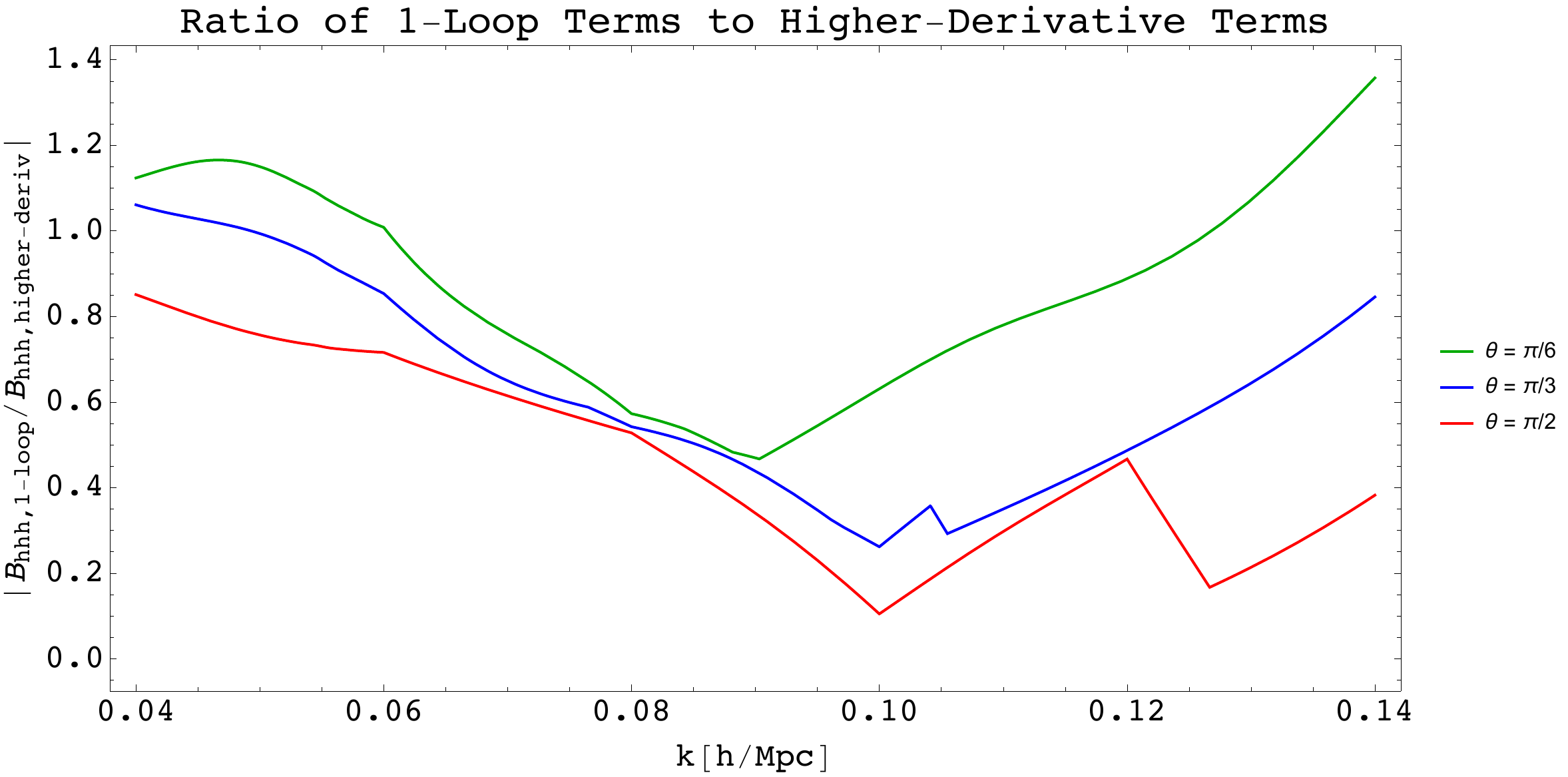} \caption{Ratio of the leading one-loop contributions to the halo-halo-halo bispectrum in the high-bias-scaling limit to the higher-derivative terms, evaluated using the best-fit bias coefficients from our Bin 3 fit. The various lines correspond to different triangular configurations $(\mathbf{k}_1, \mathbf{k}_2, \mathbf{k}_3)$, where $\lvert\mathbf{k}_1\rvert = \lvert\mathbf{k}_2\rvert$ and $\theta$ is the angle between $\mathbf{k}_1$ and $\mathbf{k}_2$. Squeezed configurations ($\theta = \pi/6$) are shown in green, equilateral configurations ($\theta = \pi/3$) are shown in blue, and rectangular configurations ($\theta = \pi/2$) are shown in red. The abrupt changes in slope result from the finite number of points at which we sample the bispectrum.}
\label{fig:ratio}
\end{figure}

\begin{figure*}[htb]
\centering
\begin{subfigure}[b]{0.4825\textwidth}
\centering
\includegraphics[width=\textwidth]{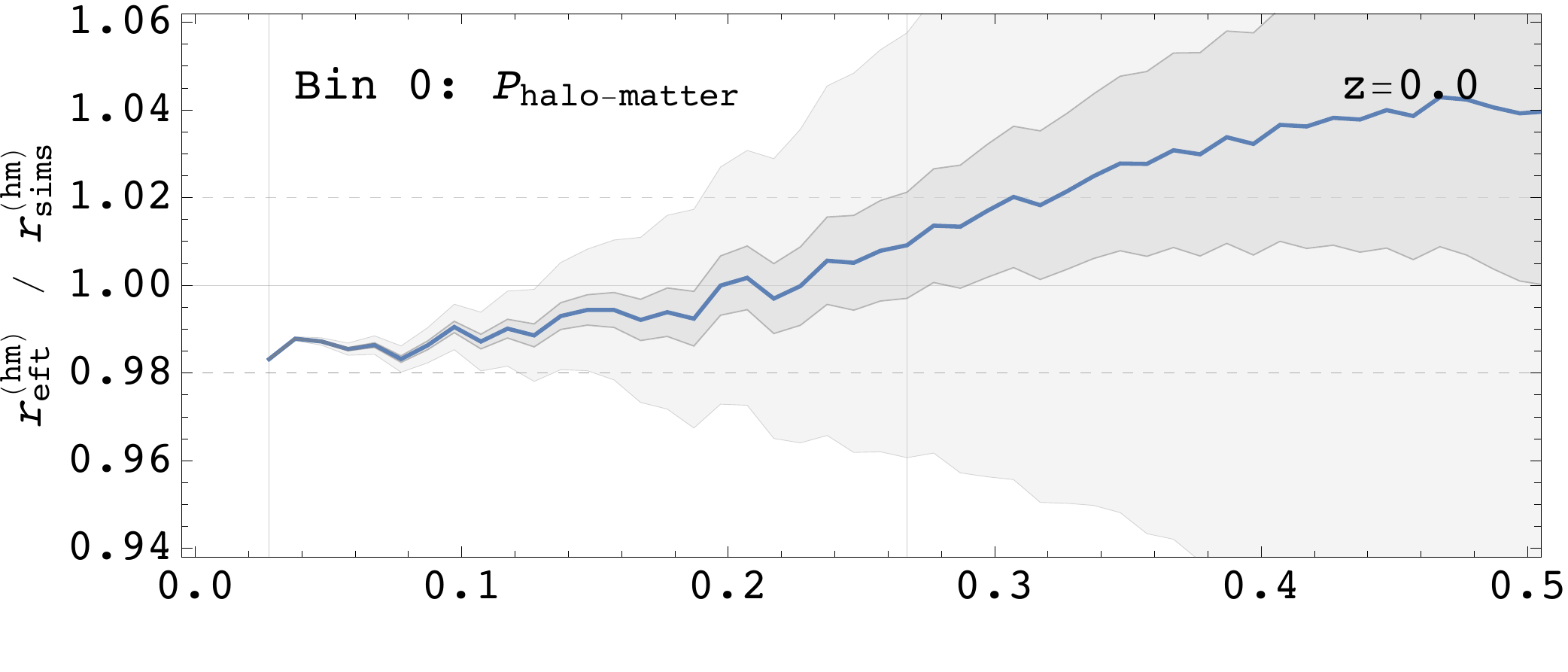}
\end{subfigure}
\hfill
\begin{subfigure}[b]{0.485\textwidth}  
\centering 
\includegraphics[width=\textwidth]{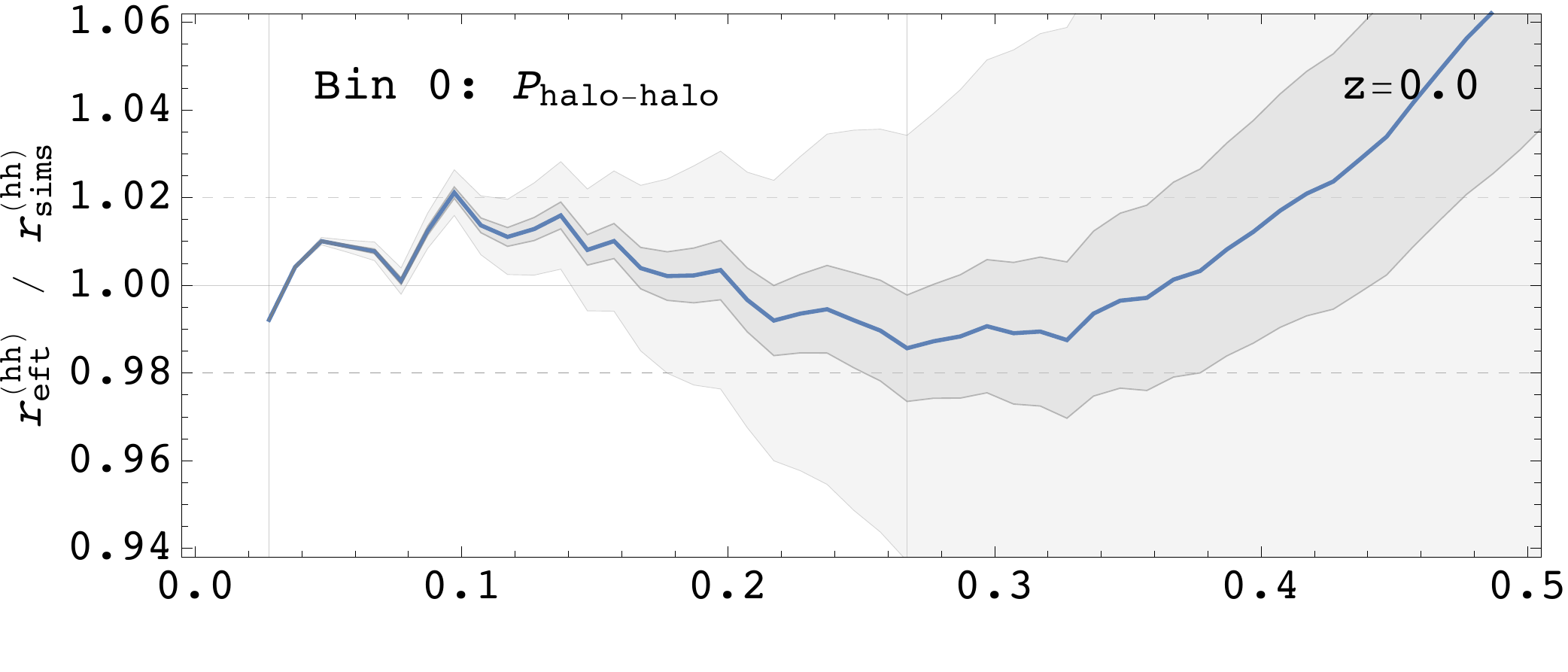}
\end{subfigure}
\begin{subfigure}[b]{0.4825\textwidth}   
\centering 
\includegraphics[width=\textwidth]{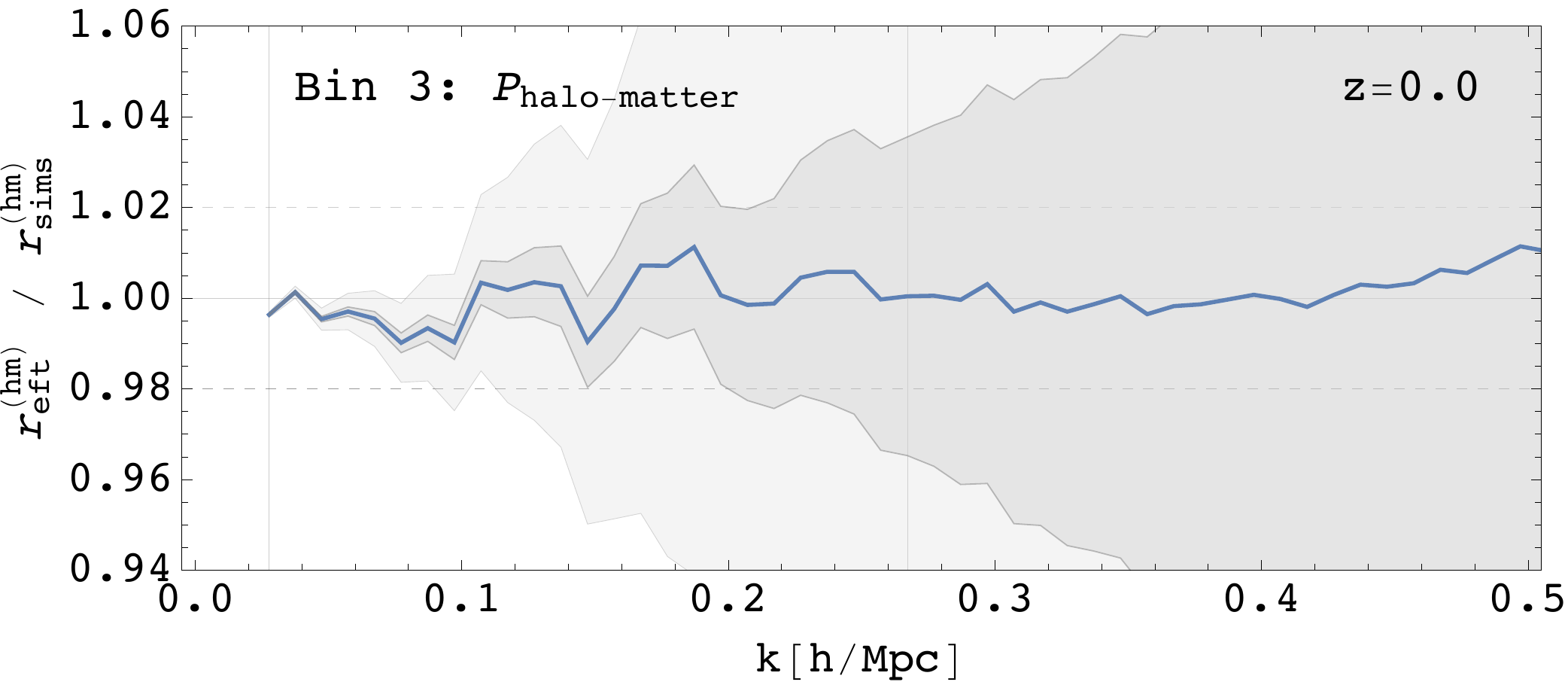}
\end{subfigure}
\quad
\begin{subfigure}[b]{0.485\textwidth}   
\centering 
\includegraphics[width=\textwidth]{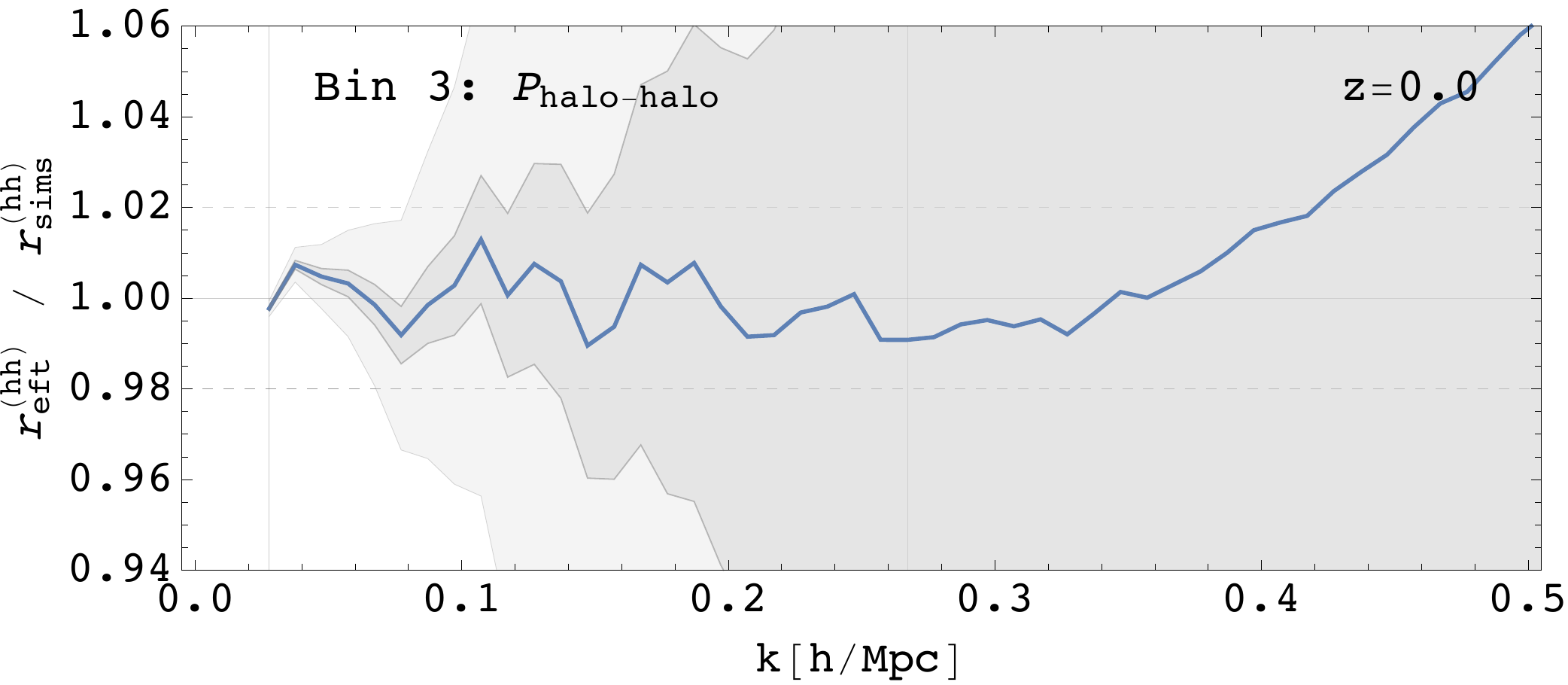}
\end{subfigure}
\caption{Ratio of the EFTofLSS predictions for the power spectra of biased tracers to results from the Millennium-XXL N-body simulation for the cross power spectrum (left column) and auto power spectrum (right column) of dark matter halos. We plot the ratio for light tracers (top row) and for very massive tracers (bottom row) using the best-fit bias coefficients from our fits that include the leading one-loop contributions to the bispectra in the high-bias-scaling limit. The shaded areas represent order-of-magnitude error estimates.} 
\label{fig:spectra}
\end{figure*}

To further assess the importance of the leading one-loop contributions to the bispectra for very massive tracers in the high-bias-scaling limit, we examine the ratio $B_{hhh,\rm{1-loop}}/B_{hhh,\rm {deriv}}$ for our Bin 3 fit that includes both higher-derivative terms and these one-loop contributions. Here, $B_{hhh,\rm{deriv}}$ is the sum of the higher-derivative terms that contribute to $B_{hhh}$. We evaluate the one-loop and higher-derivative terms using the best-fit bias coefficients at $k_{\text{fit}} = 0.15\ h\ \rm{Mpc}^{-1}$, and Figure \ref{fig:ratio} shows the ratio of these contributions for several $(\mathbf{k}_1, \mathbf{k}_2, \mathbf{k}_3)$ configurations. Figure \ref{fig:ratio} suggests that the leading one-loop contributions in the high-bias-scaling limit are generally small relative to the higher derivative terms except at the highest scales of interest\footnote{Recall that data for the three-point functions are only available up to $k = 0.14\ h\ \rm{Mpc}^{-1}$ for Bin 3.}. Nonetheless, these one-loop contributions are not completely negligible relative to the higher-derivative terms, which explains why they improve the $k$-reach of the tree-level plus higher-derivative prediction for the bispectra. Finally, Figure \ref{fig:spectra} shows the cross and auto power spectra predicted by our Bin 0 and Bin 3 fits that include the one-loop contributions derived above divided by the corresponding power spectra from the Millennium simulation. These predictions fit the corresponding power spectra at the percent level up to $k \simeq 0.2\ h\ \rm{Mpc}^{-1}$.

%----------------------------------------------------------------------------------------
%	SECTION 4
%----------------------------------------------------------------------------------------

\section{Conclusions}
\label{sec:conclusion}

In this paper, we explored the EFTofLSS prediction for the bispectra of biased tracers. In particular, we argued that the current tree-level plus higher-derivative prediction for the bispectra of very massive tracers is potentially inconsistent because certain one-loop contributions could be comparable to higher-derivative biases if higher-order bias coefficients are strongly enhanced with respect to the linear bias for such tracers. We showed that it is only necessary to compute a small, easy-to-compute subset of the full one-loop bispectra in this high-bias-scaling limit in order to check the consistency of the tree-level plus higher-derivative prediction, and we derived expressions for the relevant diagrams. Including these contributions in a fit with the power spectra and bispectra of biased tracers from a numerical simulation increases the perturbative reach of the EFTofLSS prediction by about $0.02\ h\ \rm{Mpc}^{-1}$ for very massive tracers. In particular, the $k$-reach of the fit with the one-loop power spectra and tree-level plus higher-derivative bispectra of very massive tracers improves from $k\simeq 0.17\ h\ \rm{Mpc}^{-1}$ to $k\simeq 0.19\ h\ \rm{Mpc}^{-1}$ when the leading one-loop contributions in the high-bias-scaling limit are included, which is only slightly larger than the corresponding improvement for light tracers. This relatively mild improvement suggests that these one-loop contributions are not strongly bias enhanced, and the best-fit values of the bias coefficients that we find for our Bin 3 fit are consistent with this interpretation: we do not find strong evidence that $b_{3}\gg b_{2}\gg b_{1}$. Moreover, we find that the one-loop contributions derived above are generally small relative to higher-derivative terms except at the highest scales of interest. We note that our calculation is designed only to check whether it is necessary to include one-loop contributions alongside higher-derivative contributions for very massive tracers, so the values of the bias parameters listed above should be regarded as rough estimates of the values that would be obtained by a calculation that consistently includes the complete one-loop bispectra. However, our results strongly suggest that the seven- (for light tracers) to twelve- (for massive tracers) parameter predictions for the power spectra and bispectra derived in \cite{Fujita:2016dne}, which include only higher-derivative corrections to the tree-level bispectra, fit the two- and three-point functions for biased tracers up to $k\simeq 0.17\ h\ \rm{Mpc}^{-1}$ without making spurious assumptions about the scaling of the bias coefficients or the size of the one-loop contributions. Thus, the EFTofLSS prediction for the one-loop power spectra and tree-level plus higher-derivative bispectra accurately describes the two- and three-point functions of biased tracers within a few percent up to $k\simeq 0.17\ h\ \rm{Mpc}^{-1}$ for collapsed objects of a very wide range of masses.

\begin{table}[b]
\begin{center}
\begin{tabular}{ |P{0.6cm}||P{0.6cm}|P{0.6cm}||P{0.6cm}||P{0.6cm}||P{0.6cm}||P{0.6cm}||P{0.6cm}||P{0.6cm}||P{0.6cm}||P{0.6cm}||P{0.6cm}| }
\hline
\multicolumn{12}{|c|}{Bin 0 Correlation Matrix} \\
\hline
&$b_{\delta,1}$&$b_{\delta,2}$&$b_{\delta^2}$&$b_{c_{s}}$&$b_{\epsilon\delta}$&$b_{\epsilon}$&$b_{\delta^3}$&$b_{s^3}$&$b_{st}$&$b_{\psi}$&$b_{\delta s^2}$\\
\hline
$b_{\delta,1}$   &\footnotesize{$1.$}&\footnotesize{$\text{-}0.05$}&\footnotesize{$0.15$}&\footnotesize{$\text{-}0.27$}&\footnotesize{$0.19$}&\footnotesize{$\text{-}0.07$}&\footnotesize{$\text{-}0.26$}&\footnotesize{$0.42$}&\footnotesize{$\text{-}0.56$}&\footnotesize{$0.12$}&\footnotesize{$0.25$}\\
$b_{\delta,2}$   &\footnotesize{$\text{-}0.05$}&\footnotesize{$1.$}&\footnotesize{$\text{-}0.97$}&\footnotesize{$\text{-}0.32$}&\footnotesize{$\text{-}0.09$}&\footnotesize{$\text{-}0.30$}&\footnotesize{$0.07$}&\footnotesize{$0.11$}&\footnotesize{$0.02$}&\footnotesize{$0.07$}&\footnotesize{$\text{-}0.05$}\\
$b_{\delta^2}$   &\footnotesize{$0.15$}&\footnotesize{$\text{-}0.97$}&\footnotesize{$1.$}&\footnotesize{$0.45$}&\footnotesize{$0.16$}&\footnotesize{$0.08$}&\footnotesize{$\text{-}0.15$}&\footnotesize{$\text{-}0.04$}&\footnotesize{$\text{-}0.04$}&\footnotesize{$\text{-}0.05$}&\footnotesize{$0.13$}\\
$b_{c_{s}}$   &\footnotesize{$\text{-}0.27$}&\footnotesize{$\text{-}0.32$}&\footnotesize{$0.45$}&\footnotesize{$1.$}&\footnotesize{$0.14$}&\footnotesize{$\text{-}0.54$}&\footnotesize{$\text{-}0.13$}&\footnotesize{$\text{-}0.14$}&\footnotesize{$0.48$}&\footnotesize{$\text{-}0.03$}&\footnotesize{$0.11$}\\
$b_{\epsilon\delta}$   &\footnotesize{$0.19$}&\footnotesize{$\text{-}0.09$}&\footnotesize{$0.16$}&\footnotesize{$0.14$}&\footnotesize{$1.$}&\footnotesize{$\text{-}0.21$}&\footnotesize{$\text{-}0.99$}&\footnotesize{$\text{-}0.11$}&\footnotesize{$\text{-}0.68$}&\footnotesize{$0.89$}&\footnotesize{$0.99$}\\
$b_{\epsilon}$   &\footnotesize{$\text{-}0.07$}&\footnotesize{$\text{-}0.30$}&\footnotesize{$0.08$}&\footnotesize{$\text{-}0.54$}&\footnotesize{$\text{-}0.21$}&\footnotesize{$1.$}&\footnotesize{$0.23$}&\footnotesize{$\text{-}0.18$}&\footnotesize{$\text{-}0.13$}&\footnotesize{$\text{-}0.08$}&\footnotesize{$\text{-}0.22$}\\
$b_{\delta^3}$   &\footnotesize{$\text{-}0.26$}&\footnotesize{$0.07$}&\footnotesize{$\text{-}0.15$}&\footnotesize{$\text{-}0.13$}&\footnotesize{$\text{-}0.99$}&\footnotesize{$0.23$}&\footnotesize{$1.$}&\footnotesize{$0.08$}&\footnotesize{$0.72$}&\footnotesize{$\text{-}0.90$}&\footnotesize{$\text{-}1.00$}\\
$b_{s^3}$   &\footnotesize{$0.42$}&\footnotesize{$0.11$}&\footnotesize{$\text{-}0.04$}&\footnotesize{$\text{-}0.14$}&\footnotesize{$\text{-}0.11$}&\footnotesize{$\text{-}0.18$}&\footnotesize{$0.08$}&\footnotesize{$1.$}&\footnotesize{$\text{-}0.07$}&\footnotesize{$\text{-}0.21$}&\footnotesize{$\text{-}0.08$}\\
$b_{st}$   &\footnotesize{$\text{-}0.56$}&\footnotesize{$0.02$}&\footnotesize{$\text{-}0.04$}&\footnotesize{$0.48$}&\footnotesize{$\text{-}0.68$}&\footnotesize{$\text{-}0.13$}&\footnotesize{$0.72$}&\footnotesize{$\text{-}0.07$}&\footnotesize{$1.$}&\footnotesize{$\text{-}0.76$}&\footnotesize{$\text{-}0.72$}\\
$b_{\psi}$   &\footnotesize{$0.12$}&\footnotesize{$0.07$}&\footnotesize{$\text{-}0.05$}&\footnotesize{$\text{-}0.03$}&\footnotesize{$0.89$}&\footnotesize{$\text{-}0.08$}&\footnotesize{$\text{-}0.90$}&\footnotesize{$\text{-}0.21$}&\footnotesize{$\text{-}0.76$}&\footnotesize{$1.$}&\footnotesize{$0.91$}\\
$b_{\delta s^2}$   &\footnotesize{$0.25$}&\footnotesize{$\text{-}0.05$}&\footnotesize{$0.13$}&\footnotesize{$0.11$}&\footnotesize{$0.99$}&\footnotesize{$\text{-}0.22$}&\footnotesize{$\text{-}1.00$}&\footnotesize{$\text{-}0.08$}&\footnotesize{$\text{-}0.72$}&\footnotesize{$0.91$}&\footnotesize{$1.$}\\
\hline
\end{tabular}
\caption{Correlation matrix for our Bin 0 fit with the one-loop power spectra and tree-level plus one-loop bispectra for light tracers in the high-bias-scaling limit. The strong correlations and anticorrelations among some of the bias coefficients associated with the leading one-loop contributions indicates that certain terms are not well constrained by the Bin 0 data.}\label{tab:Bin0correlation}
\end{center}
\end{table}

\begin{table}[tb]
\begin{widepage}
\begin{tabular}{ |P{0.7cm}||P{0.65cm}|P{0.6cm}||P{0.6cm}||P{0.6cm}||P{0.6cm}||P{0.6cm}||P{0.6cm}||P{0.6cm}||P{0.6cm}||P{0.6cm}||P{0.6cm}||P{0.65cm}||P{0.65cm}||P{0.6cm}||P{0.65cm}||P{0.65cm}| }
\hline
\multicolumn{17}{|c|}{Bin 3 Correlation Matrix} \\
\hline
&$b_{\delta,1}$&$b_{\delta,2}$&$b_{\delta^2}$&$b_{c_s}$&$b_{\epsilon\delta}$&$b_\epsilon$&$b_{\delta^3}$&$b_{s^3}$&$b_{st}$&$b_{\psi}$&$b_{\delta s^2}$&$b_{\partial^2\delta^{2}}$&$b_{(\partial\delta)^{2}}$&$b_{\partial^2\epsilon}$&$b_{\partial^2\epsilon\delta}$&$b_{\epsilon\partial^2\delta}$\\
\hline
$b_{\delta,1}$   &\footnotesize{$1.$}&\footnotesize{$\text{-}0.14$}&\footnotesize{$0.27$}&\footnotesize{$0.15$}&\footnotesize{$0.02$}&\footnotesize{$\text{-}0.43$}&\footnotesize{$\text{-}0.26$}&\footnotesize{$\text{-}0.06$}&\footnotesize{$\text{-}0.51$}&\footnotesize{$0.20$}&\footnotesize{$0.28$}&\footnotesize{$\text{-}0.08$}&\footnotesize{$0.04$}&\footnotesize{$\text{-}0.24$}&\footnotesize{$0.31$}&\footnotesize{$\text{-}0.04$}\\
$b_{\delta,2}$   &\footnotesize{$\text{-}0.14$}&\footnotesize{$1.$}&\footnotesize{$\text{-}0.94$}&\footnotesize{$0.10$}&\footnotesize{$\text{-}0.73$}&\footnotesize{$0.40$}&\footnotesize{$0.40$}&\footnotesize{$0.77$}&\footnotesize{$0.14$}&\footnotesize{$\text{-}0.04$}&\footnotesize{$\text{-}0.52$}&\footnotesize{$\text{-}0.47$}&\footnotesize{$0.67$}&\footnotesize{$\text{-}0.20$}&\footnotesize{$\text{-}0.29$}&\footnotesize{$0.59$}\\
$b_{\delta^2}$   &\footnotesize{$0.27$}&\footnotesize{$\text{-}0.94$}&\footnotesize{$1.$}&\footnotesize{$0.21$}&\footnotesize{$0.66$}&\footnotesize{$\text{-}0.49$}&\footnotesize{$\text{-}0.49$}&\footnotesize{$\text{-}0.78$}&\footnotesize{$\text{-}0.34$}&\footnotesize{$0.17$}&\footnotesize{$0.61$}&\footnotesize{$0.29$}&\footnotesize{$\text{-}0.54$}&\footnotesize{$0.06$}&\footnotesize{$0.48$}&\footnotesize{$\text{-}0.53$}\\
$b_{c_s}$   &\footnotesize{$0.15$}&\footnotesize{$0.10$}&\footnotesize{$0.21$}&\footnotesize{$1.$}&\footnotesize{$\text{-}0.13$}&\footnotesize{$\text{-}0.33$}&\footnotesize{$\text{-}0.22$}&\footnotesize{\text{-}$0.09$}&\footnotesize{\text{-}$0.40$}&\footnotesize{$0.34$}&\footnotesize{$0.24$}&\footnotesize{\text{-}$0.53$}&\footnotesize{$0.32$}&\footnotesize{$\text{-}0.47$}&\footnotesize{$0.51$}&\footnotesize{$0.14$}\\
$b_{\epsilon\delta}$   &\footnotesize{$0.02$}&\footnotesize{$\text{-}0.73$}&\footnotesize{$0.66$}&\footnotesize{\text{-}$0.13$}&\footnotesize{$1.$}&\footnotesize{$\text{-}0.34$}&\footnotesize{\text{-}$0.63$}&\footnotesize{$\text{-}0.63$}&\footnotesize{$\text{-}0.33$}&\footnotesize{$0.42$}&\footnotesize{$0.68$}&\footnotesize{$0.26$}&\footnotesize{\text{-}$0.70$}&\footnotesize{$0.10$}&\footnotesize{$0.12$}&\footnotesize{\text{-}$0.66$}\\
$b_\epsilon$   &\footnotesize{\text{-}$0.43$}&\footnotesize{$0.40$}&\footnotesize{$\text{-}0.49$}&\footnotesize{\text{-}$0.33$}&\footnotesize{$\text{-}0.34$}&\footnotesize{$1.$}&\footnotesize{$0.15$}&\footnotesize{$0.36$}&\footnotesize{$0.25$}&\footnotesize{\text{-}$0.17$}&\footnotesize{$\text{-}0.22$}&\footnotesize{$0.15$}&\footnotesize{$0.05$}&\footnotesize{$0.69$}&\footnotesize{\text{-}$0.33$}&\footnotesize{$0.06$}\\
$b_{\delta^3}$   &\footnotesize{$\text{-}0.26$}&\footnotesize{$0.40$}&\footnotesize{$\text{-}0.49$}&\footnotesize{$\text{-}0.22$}&\footnotesize{\text{-}$0.63$}&\footnotesize{$0.15$}&\footnotesize{$1.$}&\footnotesize{$0.47$}&\footnotesize{$0.76$}&\footnotesize{$\text{-}0.76$}&\footnotesize{\text{-}$0.99$}&\footnotesize{\text{-}$0.10$}&\footnotesize{$0.57$}&\footnotesize{\text{-}$0.07$}&\footnotesize{$\text{-}0.51$}&\footnotesize{$0.54$}\\
$b_{s^3}$   &\footnotesize{\text{-}$0.06$}&\footnotesize{$0.77$}&\footnotesize{\text{-}$0.78$}&\footnotesize{\text{-}$0.09$}&\footnotesize{$\text{-}0.63$}&\footnotesize{$0.36$}&\footnotesize{$0.47$}&\footnotesize{$1.$}&\footnotesize{$0.34$}&\footnotesize{$\text{-}0.34$}&\footnotesize{\text{-}$0.56$}&\footnotesize{$\text{-}0.16$}&\footnotesize{$0.56$}&\footnotesize{\text{-}$0.09$}&\footnotesize{\text{-}$0.46$}&\footnotesize{$0.52$}\\
$b_{st}$   &\footnotesize{\text{-}$0.51$}&\footnotesize{$0.14$}&\footnotesize{\text{-}$0.34$}&\footnotesize{\text{-}$0.40$}&\footnotesize{$\text{-}0.33$}&\footnotesize{$0.25$}&\footnotesize{$0.76$}&\footnotesize{$0.34$}&\footnotesize{$1.$}&\footnotesize{\text{-}$0.88$}&\footnotesize{$\text{-}0.75$}&\footnotesize{$0.36$}&\footnotesize{$0.10$}&\footnotesize{$0.15$}&\footnotesize{\text{-}$0.60$}&\footnotesize{$0.23$}\\
$b_{\psi}$   &\footnotesize{$0.20$}&\footnotesize{\text{-}$0.04$}&\footnotesize{$0.17$}&\footnotesize{$0.34$}&\footnotesize{$0.42$}&\footnotesize{\text{-}$0.17$}&\footnotesize{$\text{-}0.76$}&\footnotesize{$\text{-}0.34$}&\footnotesize{\text{-}$0.88$}&\footnotesize{$1.$}&\footnotesize{$0.72$}&\footnotesize{\text{-}$0.39$}&\footnotesize{$\text{-}0.15$}&\footnotesize{\text{-}$0.15$}&\footnotesize{$0.48$}&\footnotesize{$\text{-}0.24$}\\
$b_{\delta s^2}$   &\footnotesize{$0.28$}&\footnotesize{$\text{-}0.52$}&\footnotesize{$0.61$}&\footnotesize{$0.24$}&\footnotesize{$0.68$}&\footnotesize{$\text{-}0.22$}&\footnotesize{\text{-}$0.99$}&\footnotesize{\text{-}$0.56$}&\footnotesize{$\text{-}0.75$}&\footnotesize{$0.72$}&\footnotesize{$1.$}&\footnotesize{$0.14$}&\footnotesize{\text{-}$0.61$}&\footnotesize{$0.08$}&\footnotesize{$0.55$}&\footnotesize{\text{-}$0.57$}\\
$b_{\partial^2\delta^{2}}$   &\footnotesize{\text{-}$0.08$}&\footnotesize{$\text{-}0.47$}&\footnotesize{$0.29$}&\footnotesize{\text{-}$0.53$}&\footnotesize{$0.26$}&\footnotesize{$0.15$}&\footnotesize{\text{-}$0.10$}&\footnotesize{$\text{-}0.16$}&\footnotesize{$0.36$}&\footnotesize{\text{-}$0.39$}&\footnotesize{$0.14$}&\footnotesize{$1.$}&\footnotesize{$\text{-}0.41$}&\footnotesize{$0.45$}&\footnotesize{\text{-}$0.36$}&\footnotesize{$\text{-}0.23$}\\
$b_{(\partial\delta)^{2}}$   &\footnotesize{$0.04$}&\footnotesize{$0.67$}&\footnotesize{\text{-}$0.54$}&\footnotesize{$0.32$}&\footnotesize{\text{-}$0.70$}&\footnotesize{$0.05$}&\footnotesize{$0.57$}&\footnotesize{$0.56$}&\footnotesize{$0.10$}&\footnotesize{$\text{-}0.15$}&\footnotesize{\text{-}$0.61$}&\footnotesize{$\text{-}0.41$}&\footnotesize{$1.$}&\footnotesize{\text{-}$0.35$}&\footnotesize{\text{-}$0.27$}&\footnotesize{$0.79$}\\
$b_{\partial^2\epsilon}$   &\footnotesize{\text{-}$0.24$}&\footnotesize{\text{-}$0.20$}&\footnotesize{$0.06$}&\footnotesize{$\text{-}0.47$}&\footnotesize{$0.10$}&\footnotesize{$0.69$}&\footnotesize{\text{-}$0.07$}&\footnotesize{\text{-}$0.09$}&\footnotesize{$0.15$}&\footnotesize{\text{-}$0.15$}&\footnotesize{$0.08$}&\footnotesize{$0.45$}&\footnotesize{\text{-}$0.35$}&\footnotesize{$1.$}&\footnotesize{\text{-}$0.18$}&\footnotesize{\text{-}$0.29$}\\
$b_{\partial^2\epsilon\delta}$   &\footnotesize{$0.31$}&\footnotesize{\text{-}$0.29$}&\footnotesize{$0.48$}&\footnotesize{$0.51$}&\footnotesize{$0.12$}&\footnotesize{\text{-}$0.33$}&\footnotesize{$\text{-}0.51$}&\footnotesize{\text{-}$0.46$}&\footnotesize{\text{-}$0.60$}&\footnotesize{$0.48$}&\footnotesize{$0.55$}&\footnotesize{\text{-}$0.36$}&\footnotesize{\text{-}$0.27$}&\footnotesize{\text{-}$0.18$}&\footnotesize{$1.$}&\footnotesize{\text{-}$0.49$}\\
$b_{\epsilon\partial^2\delta}$   &\footnotesize{$\text{-}0.04$}&\footnotesize{$0.59$}&\footnotesize{\text{-}$0.53$}&\footnotesize{$0.14$}&\footnotesize{\text{-}$0.66$}&\footnotesize{$0.06$}&\footnotesize{$0.54$}&\footnotesize{$0.52$}&\footnotesize{$0.23$}&\footnotesize{$\text{-}0.24$}&\footnotesize{\text{-}$0.57$}&\footnotesize{$\text{-}0.23$}&\footnotesize{$0.79$}&\footnotesize{\text{-}$0.29$}&\footnotesize{\text{-}$0.49$}&\footnotesize{$1.$}\\
\hline
\end{tabular}
\caption{Correlation matrix for our Bin 3 fit with the one-loop power spectra and tree-level plus higher-derivative plus one-loop bispectra for very massive tracers in the high-bias-scaling limit. Note that we have suppressed the factors of $1/k_{M}^2$ that multiply the higher-derivative bias coefficients.}\label{tab:Bin3correlation}
\end{widepage}
\end{table}

%%%%%%%%%%%%%%%%
%
%
%        Acknowledgements
%
%
%%%%%%%%%%%%%%%%

\subsubsection*{Acknowledgments}

We thank Valentin Mauerhofer for help with the Mathematica fitting procedure. L.S.\ is partially supported by NSF award 1720397.

%%%%%%%%%%%%%%%%
%
%
%           Appendix
%
%
%%%%%%%%%%%%%%%%

\begin{appendix}\appendix

\section{IR-Safe Integrands}
\label{sec:IR-safe}

Single loop diagrams in the EFTofLSS generically contain IR-divergences that, for IR-safe observables, cancel in the final result. However, eliminating these divergences from each diagram can improve the accuracy of the calculations. The leading one-loop contributions to the bispectra for biased tracers in the high-bias-scaling limit contain such IR-divergences, and we follow the treatment of IR-safety for the one-loop matter bispectrum in \cite{Angulo:2014tfa} to eliminate these contributions. 

To make the $B_{321}$ integrand IR-safe, we follow \cite{Angulo:2014tfa} by mapping each potential IR-divergence to $\mathbf{q} = \mathbf{0}$, where $\mathbf{q}$ is the integration variable. In \eqn{eq:b321eq}, note that the integrand of $B_{321}$, which we denote as $b_{321}(\mathbf{q})$, is symmetric under $\mathbf{q} \rightarrow \mathbf{k}_{1}-\mathbf{q}$ if we symmetrize the kernels. This implies
\begin{equation} B_{321} = \int_{\mathbf{q}} b_{321}(\mathbf{q}) = 2\int_{\mathbf{q}} b_{321}(\mathbf{q})\Theta(\lvert\mathbf{k}_{1}-\mathbf{q}\rvert-q), \end{equation}
with an appropriate $\Theta$-function for each permutation, where we have suppressed the dependence of the external momenta and $\int_{\mathbf{q}}\equiv \int \frac{\text{d}^{3}q}{(2\pi)^3}$.

To make the $B_{222}$ integrand IR-safe, we again follow \cite{Angulo:2014tfa} by multiplying the integrand $b_{222}(\mathbf{q})$ by a product of $\Theta$-functions to map each potential divergence to $\mathbf{q}=\mathbf{0}$. $B_{222}$ then takes the form
\begin{equation} B_{222} = \int_{\mathbf{q}}\big[b_{222}^{(k_{3}>k_{1})}(\mathbf{q})\Theta(k_{3}-k_{1}) + b_{222}^{(k_{1}>k_{3})}(\mathbf{q})\Theta(k_{1}-k_{3})\big], \end{equation}
where the $b_{222}^{(k_{1}>k_{3})}$ terms are the same as those provided in \cite{Angulo:2014tfa}, with the replacement $F_{s}^{(2)} \rightarrow \tilde{K}_{s}^{(2)}$. %\footnote{Note that in Eq.\ (16) of \cite{Angulo:2014tfa}, the last kernel in the second term should read $F_{2}^{(s)}(-\mathbf{k}_{1}+\mathbf{\widetilde{q}}, -\mathbf{k}_{3} - \mathbf{\widetilde{q}})$.} 
The IR-divergences cancel without $B_{411}$, since $B_{411}$ is degenerate with the tree-level bispectra in the high-bias-scaling limit. We find that the IR-safe integrals take longer to compute than their non-IR-safe counterparts and that they do not significantly affect our numerical results. Thus, we do not implement this procedure for the one-loop contributions that enter our final calculations.

\end{appendix}

\bibliography{references.bib}

\providecommand{\href}[2]{#2}\begingroup\raggedright\begin{thebibliography}{10}

\bibitem{2013arXiv1308.0847L}
M.~{Levi \emph{et al.} [DESI Collaboration]}, {\it {The DESI Experiment, a
  whitepaper for Snowmass 2013}},  {\em ArXiv e-prints} (2013)
  [\href{http://arxiv.org/abs/1308.0847}{{\tt arXiv:1308.0847}}].

\bibitem{2009arXiv0912.0201L}
P.~A. {Abell \emph{et al.} [LSST Science and LSST Project Collaborations]},
  {\it {LSST Science Book, Version 2.0}},  {\em ArXiv e-prints} (2009)
  [\href{http://arxiv.org/abs/0912.0201}{{\tt arXiv:0912.0201}}].

\bibitem{Baumann:2010tm}
D.~Baumann, A.~Nicolis, L.~Senatore, and M.~Zaldarriaga, {\it {Cosmological
  Non-Linearities as an Effective Fluid}},  {\em JCAP} {\bf 1207} (2012) 051,
  [\href{http://arxiv.org/abs/1004.2488}{{\tt arXiv:1004.2488}}].

\bibitem{Carrasco:2012cv}
J.~J.~M. Carrasco, M.~P. Hertzberg, and L.~Senatore, {\it {The Effective Field
  Theory of Cosmological Large Scale Structures}},  {\em JHEP} {\bf 09} (2012)
  082, [\href{http://arxiv.org/abs/1206.2926}{{\tt arXiv:1206.2926}}].

\bibitem{Porto:2013qua}
R.~A. Porto, L.~Senatore, and M.~Zaldarriaga, {\it {The Lagrangian-space
  Effective Field Theory of Large Scale Structures}},  {\em JCAP} {\bf 1405}
  (2014) 022, [\href{http://arxiv.org/abs/1311.2168}{{\tt arXiv:1311.2168}}].

\bibitem{Senatore:2014via}
L.~Senatore and M.~Zaldarriaga, {\it {The IR-resummed Effective Field Theory of
  Large Scale Structures}},  {\em JCAP} {\bf 1502} (2015), no.~02 013,
  [\href{http://arxiv.org/abs/1404.5954}{{\tt arXiv:1404.5954}}].

\bibitem{Carrasco:2013sva}
J.~J.~M. Carrasco, S.~Foreman, D.~Green, and L.~Senatore, {\it {The 2-loop
  matter power spectrum and the IR-safe integrand}},  {\em JCAP} {\bf 1407}
  (2014) 056, [\href{http://arxiv.org/abs/1304.4946}{{\tt arXiv:1304.4946}}].

\bibitem{Carrasco:2013mua}
J.~J.~M. Carrasco, S.~Foreman, D.~Green, and L.~Senatore, {\it {The Effective
  Field Theory of Large Scale Structures at Two Loops}},  {\em JCAP} {\bf 1407}
  (2014) 057, [\href{http://arxiv.org/abs/1310.0464}{{\tt arXiv:1310.0464}}].

\bibitem{Pajer:2013jj}
E.~Pajer and M.~Zaldarriaga, {\it {On the Renormalization of the Effective
  Field Theory of Large Scale Structures}},  {\em JCAP} {\bf 1308} (2013) 037,
  [\href{http://arxiv.org/abs/1301.7182}{{\tt arXiv:1301.7182}}].

\bibitem{Carroll:2013oxa}
S.~M. Carroll, S.~Leichenauer, and J.~Pollack, {\it {Consistent effective
  theory of long-wavelength cosmological perturbations}},  {\em Phys. Rev.}
  {\bf D90} (2014), no.~2 023518, [\href{http://arxiv.org/abs/1310.2920}{{\tt
  arXiv:1310.2920}}].

\bibitem{Mercolli:2013bsa}
L.~Mercolli and E.~Pajer, {\it {On the velocity in the Effective Field Theory
  of Large Scale Structures}},  {\em JCAP} {\bf 1403} (2014) 006,
  [\href{http://arxiv.org/abs/1307.3220}{{\tt arXiv:1307.3220}}].

\bibitem{Angulo:2014tfa}
R.~E. Angulo, S.~Foreman, M.~Schmittfull, and L.~Senatore, {\it {The One-Loop
  Matter Bispectrum in the Effective Field Theory of Large Scale Structures}},
  {\em JCAP} {\bf 1510} (2015) 039, [\href{http://arxiv.org/abs/1406.4143}{{\tt
  arXiv:1406.4143}}].

\bibitem{Baldauf:2014qfa}
T.~Baldauf, L.~Mercolli, M.~Mirbabayi, and E.~Pajer, {\it {The Bispectrum in
  the Effective Field Theory of Large Scale Structure}},  {\em JCAP} {\bf 1505}
  (2015), no.~05 007, [\href{http://arxiv.org/abs/1406.4135}{{\tt
  arXiv:1406.4135}}].

\bibitem{Senatore:2014eva}
L.~Senatore, {\it {Bias in the Effective Field Theory of Large Scale
  Structures}},  {\em JCAP} {\bf 1511} (2015), no.~11 007,
  [\href{http://arxiv.org/abs/1406.7843}{{\tt arXiv:1406.7843}}].

\bibitem{Senatore:2014vja}
L.~Senatore and M.~Zaldarriaga, {\it {Redshift Space Distortions in the
  Effective Field Theory of Large Scale Structures}},
  \href{http://arxiv.org/abs/1409.1225}{{\tt arXiv:1409.1225}}.

\bibitem{Lewandowski:2014rca}
M.~Lewandowski, A.~Perko, and L.~Senatore, {\it {Analytic Prediction of
  Baryonic Effects from the EFT of Large Scale Structures}},  {\em JCAP} {\bf
  1505} (2015) 019, [\href{http://arxiv.org/abs/1412.5049}{{\tt
  arXiv:1412.5049}}].

\bibitem{Mirbabayi:2014zca}
M.~Mirbabayi, F.~Schmidt, and M.~Zaldarriaga, {\it {Biased Tracers and Time
  Evolution}},  {\em JCAP} {\bf 1507} (2015), no.~07 030,
  [\href{http://arxiv.org/abs/1412.5169}{{\tt arXiv:1412.5169}}].

\bibitem{Foreman:2015uva}
S.~Foreman and L.~Senatore, {\it {The EFT of Large Scale Structures at All
  Redshifts: Analytical Predictions for Lensing}},  {\em JCAP} {\bf 1604}
  (2016) 033, [\href{http://arxiv.org/abs/1503.01775}{{\tt arXiv:1503.01775}}].

\bibitem{Angulo:2015eqa}
R.~Angulo, M.~Fasiello, L.~Senatore, and Z.~Vlah, {\it {On the Statistics of
  Biased Tracers in the Effective Field Theory of Large Scale Structures}},
  {\em JCAP} {\bf 1509} (2015) 029,
  [\href{http://arxiv.org/abs/1503.08826}{{\tt arXiv:1503.08826}}].

\bibitem{McQuinn:2015tva}
M.~McQuinn and M.~White, {\it {Cosmological perturbation theory in 1+1
  dimensions}},  {\em JCAP} {\bf 1601} (2016), no.~01 043,
  [\href{http://arxiv.org/abs/1502.07389}{{\tt arXiv:1502.07389}}].

\bibitem{Assassi:2015jqa}
V.~Assassi, D.~Baumann, E.~Pajer, Y.~Welling, and D.~van~der Woude, {\it
  {Effective theory of large-scale structure with primordial non-Gaussianity}},
   {\em JCAP} {\bf 1511} (2015) 024,
  [\href{http://arxiv.org/abs/1505.06668}{{\tt arXiv:1505.06668}}].

\bibitem{Baldauf:2015tla}
T.~Baldauf, E.~Schaan, and M.~Zaldarriaga, {\it {On the reach of perturbative
  descriptions for dark matter displacement fields}},  {\em JCAP} {\bf 1603}
  (2016), no.~03 017, [\href{http://arxiv.org/abs/1505.07098}{{\tt
  arXiv:1505.07098}}].

\bibitem{Fujita:2016dne}
T.~Fujita, V.~Mauerhofer, L.~Senatore, Z.~Vlah, and R.~Angulo, {\it {Very
  Massive Tracers and Higher Derivative Biases}},
  \href{http://arxiv.org/abs/1609.00717}{{\tt arXiv:1609.00717}}.

\bibitem{Perko:2016puo}
A.~Perko, L.~Senatore, E.~Jennings, and R.~H. Wechsler, {\it {Biased Tracers in
  Redshift Space in the EFT of Large-Scale Structure}},
  \href{http://arxiv.org/abs/1610.09321}{{\tt arXiv:1610.09321}}.

\bibitem{Lewandowski:2016yce}
M.~Lewandowski, A.~Maleknejad, and L.~Senatore, {\it {An effective description
  of dark matter and dark energy in the mildly non-linear regime}},
  \href{http://arxiv.org/abs/1611.07966}{{\tt arXiv:1611.07966}}.

\bibitem{Lewandowski:2017kes}
M.~Lewandowski and L.~Senatore, {\it {IR-safe and UV-safe integrands in the
  EFTofLSS with exact time dependence}},
  \href{http://arxiv.org/abs/1701.07012}{{\tt arXiv:1701.07012}}.

\bibitem{Senatore:2017hyk}
L.~Senatore and M.~Zaldarriaga, {\it {The Effective Field Theory of Large-Scale
  Structure in the presence of Massive Neutrinos}},
  \href{http://arxiv.org/abs/1710.04698}{{\tt arXiv:1710.04698}}.

\bibitem{Senatore:2017pbn}
L.~Senatore and G.~Trevisan, {\it {On the IR-Resummation in the EFTofLSS}},
  \href{http://arxiv.org/abs/1710.02178}{{\tt arXiv:1710.02178}}.

\bibitem{Angulo12033216}
R.~E. {Angulo \emph{et al.}}, {\it {Scaling relations for galaxy clusters in
  the Millennium-XXL simulation}},  {\em \mnras} {\bf 426} (Nov., 2012)
  2046--2062, [\href{http://arxiv.org/abs/1203.3216}{{\tt arXiv:1203.3216}}].

\bibitem{Lazeyras:2015lgp}
T.~Lazeyras, C.~Wagner, T.~Baldauf, and F.~Schmidt, {\it {Precision measurement
  of the local bias of dark matter halos}},  {\em JCAP} {\bf 1602} (2016),
  no.~02 018, [\href{http://arxiv.org/abs/1511.01096}{{\tt arXiv:1511.01096}}].

\bibitem{Bernardeau2002}
F.~{Bernardeau}, S.~{Colombi}, E.~{Gazta{\~n}aga}, and R.~{Scoccimarro}, {\it
  {Large-scale structure of the Universe and cosmological perturbation
  theory}},  {\em Phys. Rept.} {\bf 367} (Sept., 2002) 1--248,
  [\href{http://arxiv.org/abs/astro-ph/0112551}{{\tt astro-ph/0112551}}].

\end{thebibliography}\endgroup

\end{document}